\documentclass[12pt]{article}
\usepackage{epsfig}

\newcommand{\be}{\begin{equation}}
\newcommand{\ee}{\end{equation}}

\def\bsg{\ifmmode B\to X_s\gamma\else $B\to X_s\gamma$\fi}
\def\bsll{\ifmmode B\to X_s\ell^+\ell^-\else $B\to X_s\ell^+\ell^-$\fi}
\def\shat{\ifmmode \hat{s}\else $\hat{s}$\fi}

\newcommand{\newc}{\newcommand}

\newc{\gsim}{\lower.7ex\hbox{$\;\stackrel{\textstyle>}{\sim}\;$}}
\newc{\lsim}{\lower.7ex\hbox{$\;\stackrel{\textstyle<}{\sim}\;$}}
\newc{\ie}{{\it i.e.}}
\newc{\etal}{{\it et al.}}
\newc{\mev}{\hbox{\rm\,MeV}}
\newc{\gev}{\hbox{\rm\,GeV}}
\newc{\tev}{\hbox{\rm\,TeV}}
\newc{\xpb}{\hbox{\rm\, pb}}
\newc{\xfb}{\hbox{\rm\, fb}}

\newc{\mtop}{m_t}
\newc{\mbot}{m_b}
\newc{\mz}{M_Z}
\newc{\mw}{M_W}
\newc{\alphasmz}{\alpha_s(M_Z)}
\newc{\swsq}{\sin^2\theta_W}
\newc{\cwsq}{\cos^2\theta_W}
\newc{\tw}{\tan\theta_W}
\newc{\cw}{\cos\theta_W}
\newc{\sw}{\sin\theta_W}
\newc{\BR}{\hbox{\rm BR}}
\newc{\zbb}{Z\to b\bar}
\newc{\Gb}{\Gamma (Z\to b\bar b)}
\newc{\Gh}{\Gamma (Z\to \hbox{\rm hadrons})}
\newc{\sgn}{\mbox{sgn}}

\def\eq#1{eq.~(\ref{#1})}
\def\fig#1{fig.~\ref{#1}}

\def\vev#1{\langle {#1} \rangle}

\newlength{\myem}
\newcounter{mysubequation}[equation]

\newcommand{\GeV}{\,\mathrm{GeV}}

\def\beq{\begin{equation}}
\def\eeq{\end{equation}}
\def\bea{\begin{eqnarray}}
\def\eea{\end{eqnarray}}
\def\slashchar#1{\setbox0=\hbox{$#1$}           % set a box for #1
   \dimen0=\wd0                                 % and get its size
   \setbox1=\hbox{/} \dimen1=\wd1               % get size of /
   \ifdim\dimen0>\dimen1                        % #1 is bigger
      \rlap{\hbox to \dimen0{\hfil/\hfil}}      % so center / in box
      #1                                        % and print #1
   \else                                        % / is bigger
      \rlap{\hbox to \dimen1{\hfil$#1$\hfil}}   % so center #1
      /                                         % and print /
   \fi}                                         %
\catcode`@=11
\long\def\@caption#1[#2]#3{\par\addcontentsline{\csname
  ext@#1\endcsname}{#1}{\protect\numberline{\csname
  the#1\endcsname}{\ignorespaces #2}}\begingroup
    \small
    \@parboxrestore
    \@makecaption{\csname fnum@#1\endcsname}{\ignorespaces #3}\par
  \endgroup}
\catcode`@=12

\begin{document}

\baselineskip=18pt

\setcounter{footnote}{0}
\setcounter{figure}{0}
\setcounter{table}{0}

\begin{titlepage}
\begin{flushright}
CERN-PH-TH/2006-105
\end{flushright}
\vspace{.3in}
\begin{center}
{\Large \bf Living Dangerously with Low-Energy Supersymmetry}

\vspace{0.5cm}

{\bf Gian F. Giudice} and {\bf Riccardo
Rattazzi}

\vspace{.5cm}

{\it  CERN, Theory Division, CERN, CH-1211 Geneva 23, Switzerland}
\end{center}
\vspace{.8cm}

\begin{abstract}
\medskip
We stress that the lack of direct evidence for supersymmetry forces the soft mass
parameters to lie very close to the critical line separating the broken and unbroken 
phases of the electroweak gauge symmetry. We argue that the level of criticality, or 
fine-tuning,  that is needed to escape the present collider bounds can be quantitatively 
accounted for by assuming that the overall scale of the soft terms is an environmental 
quantity. Under fairly general assumptions, vacuum-selection considerations force a 
little hierarchy in the ratio between $m_Z^2$ and the supersymmetric particle square masses,
with a most probable value equal to a one-loop factor.
\end{abstract}

\bigskip
\bigskip

%% \begin{flushleft}
%% June 2004
%% \end{flushleft}

\end{titlepage}

%%%%%%%%%%%%%%%%%%%%%%%%%%%%%%%%%%%%%%%%%%%%%%%%%%%%%%%%%%%%%%%
%\tableofcontents
%\vfill\eject

\section{Introduction}
\label{intro}
For almost three decades, the gauge hierarchy problem has been  the only reason 
to think that the Standard Model (SM) should be overthrown right around the weak scale.
It has inspired the construction of a huge stack of new models and is arguably one of the
main motivation to build the Large Hadron Collider. As it is normally formulated,
the problem lies in the difficulty to understand the relatively low value of the Higgs mass parameter
$|m_H^2|\sim (100 \GeV)^2$ in a framework in which the SM is valid up to some ultra-high
scale $\Lambda$,  for instance for $\Lambda$ of the order of the Planck scale $M_P$.
We can equivalently picture the problem as one of criticality. Imagine the fundamental theory
at the Planck scale  has a few free parameters. In string theory, to be perhaps more concrete, 
these parameters may  correspond to  the (discrete) set of vacuum expectation values of the moduli 
fields. Let us consider the phase diagram for electroweak symmetry breaking, in the space of these
parameters.
Over the bulk  of the parameter space, $|m_H^2|$ is expected to be  of order $M_P^2$,
and therefore either $\langle H\rangle \sim M_P$ or $\langle H\rangle=0$ 
depending on the sign of $m_H^2$. The hierarchy problem is now simply stated as: 
if the critical line separating the two phases is not  special  
from the point of view of the fundamental theory, why are the parameters in the real world 
so chosen as to lie practically atop the critical line? 
%Working in Planck units, the distance from the critical line parametrizes the ration $\la
% This is to say that, by  working in Planck units
% $H\equiv M_P h$ the potential 
%  \begin{equation}
%  V/M_P^4=\epsilon h^2+\lambda h^4\simeq \lambda h^4$
%\end{equation}
%is purely quartic, ie. critical,  to an extremely good approximation $\epsilon \sim 10^{-34}$. 

Supersymmetry is relevant to this puzzle for two reasons. First, 
because it selects the critical line as a locus of enhanced symmetry\footnote{Conformal 
symmetry provides in
principle an alternative symmetry principle. The reason why it is not viable is very simple:
the presence of a fundamental physics scale, say $M_P$,  both defines the hierarchy problem and
explicitly  breaks conformal invariance.}. 
%More precisely, in the MSSM with both supersymmetry and
%PQ symmetry unbroken, the Higgs potential is indeed `critical',  $V\propto (|H_1|^2-|H_2|^2)^2$,
%at least in the directions $|H_1| \not = |H_2|$.  
More precisely, in the minimal supersymmetric SM with both supersymmetry and
Peccei-Quinn (PQ) symmetry unbroken, the Higgs potential   $V\propto (|H_1|^2-|H_2|^2)^2$ is indeed 
``critical'',
in the sense that the symmetric $H_1=H_2=0$ point  is a minimum of the potential, but it   can
be destabilized by arbitrarily small mass perturbations.
Second, supersymmetry is, under rather general
circumstances, broken only by tiny non-perturbative effects~\cite{wit}. These will  unavoidably 
move the theory slightly off the critical line.
Effectively this corresponds to the generation of tiny mass terms $\sim M_Pe^{-1/\alpha}\ll M_P$
which generically lead to electroweak symmetry breakdown  (while
stabilizing at the same time the flat direction $|H_1|  = |H_2|$) at a correspondingly low scale.
 
 In hidden sector models,  at energies below the Planck mass $M_P$,
 supersymmetry breaking is accurately parametrized by 
 soft supersymmetry-breaking terms of order $M_S$. The electroweak vacuum dynamics
 is then controlled by Renormalization-Group (RG) evolution of the soft terms from $M_P$ down
 to $M_S$. One nice feature of this evolution is that, over a wide region of the soft parameter space, 
 one of the eigenvalues of the Higgs squared-mass matrix flows to a negative value 
somewhere between $M_P$  and $M_S$~\cite{rad}. This makes 
electroweak symmetry breaking a rather natural phenomenon
 within supersymmetric extensions of the SM. 

\begin{figure}[t!]
\begin{center}
\includegraphics[width=16cm]{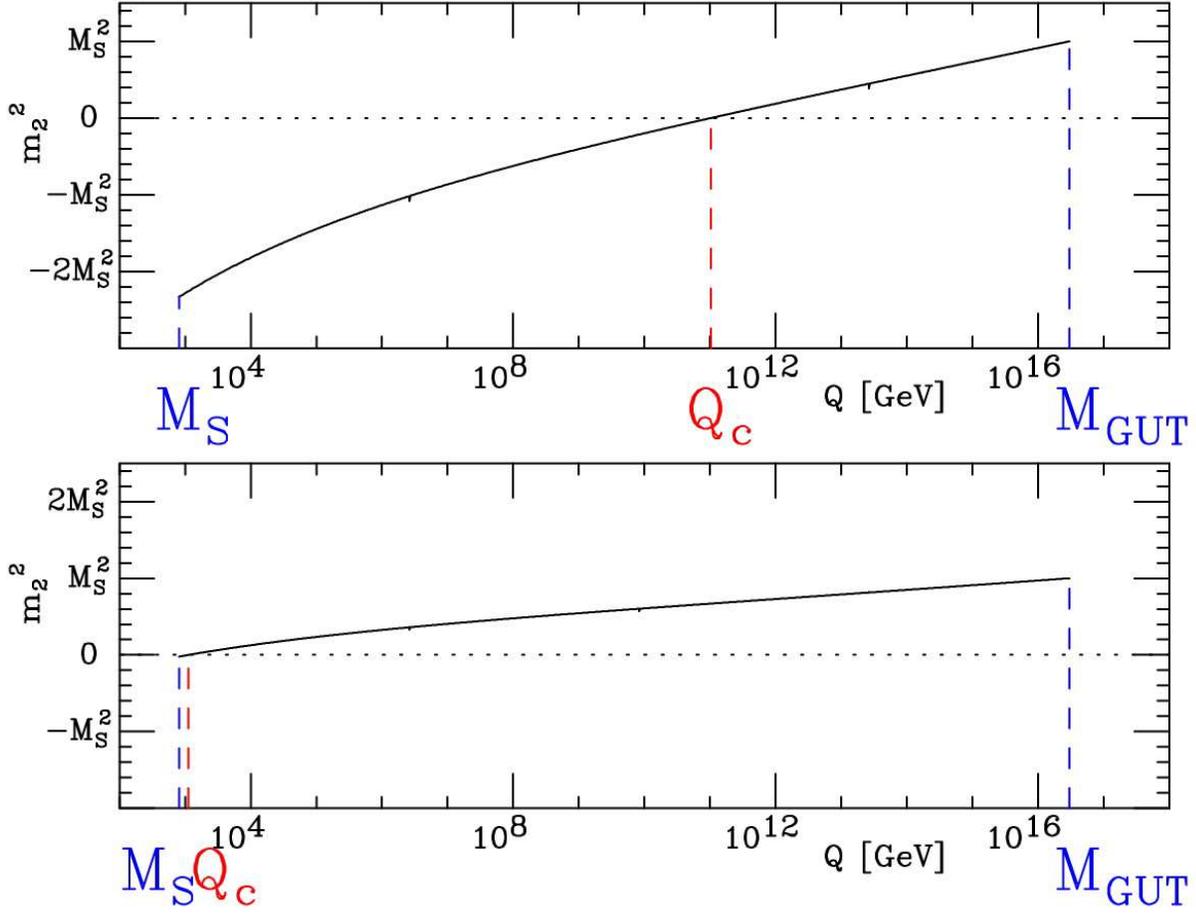}
\end{center}
\caption{ 
The running of the Higgs mass parameter $m_2^2$ as a function
of the RG scale $Q$. The top frame shows the case of a generic supersymmetric setup,
leading to $|m_2^2(M_S)|=O(M_S^2)$ and $M_S\ll Q_c\ll M_P$. The bottom frame
corresponds to a fine-tuned choice of soft terms, such that $|m_2^2(M_S)|\ll M_S^2$ and $M_S \simeq Q_c$.
 }
\label{fig5}
\end{figure}

For the sake of this discussion we should, 
however, be slightly
 more precise. Notice that,
  since the RG evolution is homogeneous in the soft terms, the RG 
scale $Q_c$ at which the Higgs mass eigenvalue crosses zero depends on $M_P$ 
and on dimensionless ratios of soft parameters, but it
is parametrically  unrelated to $M_S$.
 Furthermore,  as long as the Higgs mass matrix
 is positive definite at $M_P$,  since the evolution is logarithmic in the  RG scale,
 $Q_c$  is exponentially suppressed with respect to $M_P$ (see \fig{fig5}, top frame).   
 Therefore, the supersymmetric parameter space is essentially divided into two regions (phases) characterized respectively by $M_S\ll Q_c\ll M_P$  and $Q_c\ll M_S\ll M_P $.
 In the first region, at the scale $M_S$ where RG evolution of the soft terms
is frozen, the Higgs mass matrix has a negative eigenvalue of magnitude $\sim M_S^2$,  due to the 
 hierarchical separation between $M_S$ and $Q_c$. Given the structure and size of the 
Higgs quartic potential, this implies a weak scale $\langle 
H\rangle^2\sim M_S^2/g^2$ and $m_Z^2\sim M_S^2$. In the second region,  
RG evolution is  frozen with a positive definite Higgs mass matrix so that
the Higgs field does not break electroweak symmetry. We call this region the unbroken phase, although
 electroweak symmetry is still spontaneoulsly  broken, but only by
 fermion condensation in QCD. The resulting
 spectrum is therefore vastly different than in the $\langle H\rangle \sim  M_S/g$ phase.
 All elementary SM particles,  including $W$ and $Z$, weigh less than about 100 MeV,
while the superparners are still at $M_S$, so that the pattern is $m_Z<\Lambda_{\rm QCD}\ll M_S$. 
While the unbroken phase region does not resemble even approximately the world we live in,
the broken phase region makes instead supersymmetry relevant to phenomenology, as it solves
the gauge hierarchy problem and explains electroweak symmetry breaking in a unitary conceptual framework.

The generic spectrum $M_S\sim m_Z$ of the broken phase also held up great expectations
for a discovery of supersymmetry at LEP~\cite{natur}. 
As  those expectations were then frustrated by the experimental data,
in the post-LEP era also the broken phase does not seem to qualitatively describe our world.
The direct and indirect limits placed by LEP  point instead to a spectrum 
where $M_S$ is at least an order of magnitude larger than 
$m_Z(\gg \Lambda_{\rm QCD})$,
corresponding to the boundary between the two phases. 
The strongest, but not unique, constraint is given by  the experimental lower limit
 on the mass of the lightest CP-even Higgs $m_h$. Given 
 the tree-level theoretical upper bound $m_h<m_Z|\cos 2\beta |$,
the experimental constraint  can be satisfied only by pumping up the top-stop
quantum corrections to the Higgs quartic coupling~\cite{higcor}. 
Over most of the parameter space, this implies stop masses $m_{\tilde t}$
that range closer to a TeV than to 100 GeV. 
We can then work out where $Q_c$ should be, by expanding the RG evolution of the negative mass eigenvalue in the Higgs potential between $Q_c$ and $M_S$. For the sake of the argument we can
focus on the case $\tan\beta\to \infty$, where electroweak breaking is driven by the
Higgs mass parameter $m_2^2$ and where we find
\beq
\frac{m_Z^2}{2}=-m_2^2\simeq
\left. \frac{dm_2^2}{d\ln Q}\right|_{Q_c} \ln \frac{Q_c}{M_S}.
\eeq
%\bea
%\frac{-m_Z^2}{2}(1+\delta)= m_2^2&=&-\dot m_2^2 L_S+O(L_S^2)\\
%&=& \left [3\lambda_t^2\left (m_{\tilde t_L}^2+m_{\tilde t_R}^2+|A_t|^2\right )
%-\frac{3}{5}g_1^2 \left (M_1^2+\mu^2\right )
% -g_2^2 \left (M_2^2+\mu^2\right )\right ] L_S+O(L_S^2)
% \label{expand}
% \eea
%where $\frac{1}{8\pi^2}\ln(Q_c/M_S)=L_S $
%and  $\delta \sim 1$ is the 1-loop correction to the Higgs quartic.
For typical choices of supersymmetric parameters, the stop masses
dominate the RG evolution and $dm_2^2/d\ln Q\simeq 0.1 m_{\tilde t}^2$.
 For $m_{\tilde t}\sim 1$ TeV, we find $\ln (Q_c/M_S) <1$, which is so small that there is
not even a meaningful scale separation between the overall supersymmetric scale
$M_S$  and $Q_c$ (see \fig{fig5}, bottom frame). 
The coincidence of these two conceptually unrelated mass
scales  is one way of viewing  the fine-tuning problem of supersymmetric models.
Why should the fundamental theory prefer such {\it critical} choice of parameters?
A more quantitative illustration of this question will be given in the next section 
and the rest of the paper is an attempt to provide an answer.

\section{Fine-Tuning and Criticality in Low-Energy Supersymmetry}
\label{fine}

In this section, we want to explain, in a more quantitative fashion,
the connection between fine tuning and criticality. Let us consider
the phase diagram in the parameter space of the minimal supersymmetric
model, spanned by all independent dimensionless ratios of the 
coefficients of soft
supersymmetry-breaking terms. For illustrative purposes, we reduce 
this multi-dimensional space into a plane, by taking unified gaugino masses ($M$)
and universal scalar masses ($m^2$) at the GUT scale. For the moment, we also 
set to zero all trilinear soft terms at the GUT scale ($A=0$), and choose a
small bilinear term $B$ at the scale $M_S$, corresponding to a fixed and moderately large value
of $\tan\beta$, in the region where radiative electroweak breaking occurs. 
These hypotheses are just meant to simplify the visualization,
but the discussion we present here remains valid also for general soft terms.  
In the case under consideration, the phase diagram can be described in terms
of only two variables, which we take to be $m^2/\mu^2$ and $M^2/\mu^2$, the square ratios of
the common scalar and gaugino masses to the $\mu$ parameter, with all quantities
defined at the GUT scale. 

The SM presents two phases, with broken ($\langle H \rangle \ne 0$)
or unbroken  ($\langle H \rangle = 0$) electroweak symmetry. The situation
is more complicated in the supersymmetric version,
because of the extended structure of the Higgs sector and of the properties of
supersymmetry. We recall that the Higgs potential, along the neutral field components is
\beq
V=\frac{g^2+g^{\prime 2}}{8}\left( |H_1|^2-|H_2|^2\right)^2 +m_1^2|H_1|^2
+m_2^2|H_2|^2-m_3^2\left( H_1H_2+{\rm h.c.}\right) ,
\label{higgspot}
\eeq
and that we are working in the limit of small $m_3^2$. The boundary
condition at the GUT scale is $m_{1,2}^2=m^2+\mu^2$.

\begin{figure}[t!]
\begin{center}
\includegraphics[width=16cm]{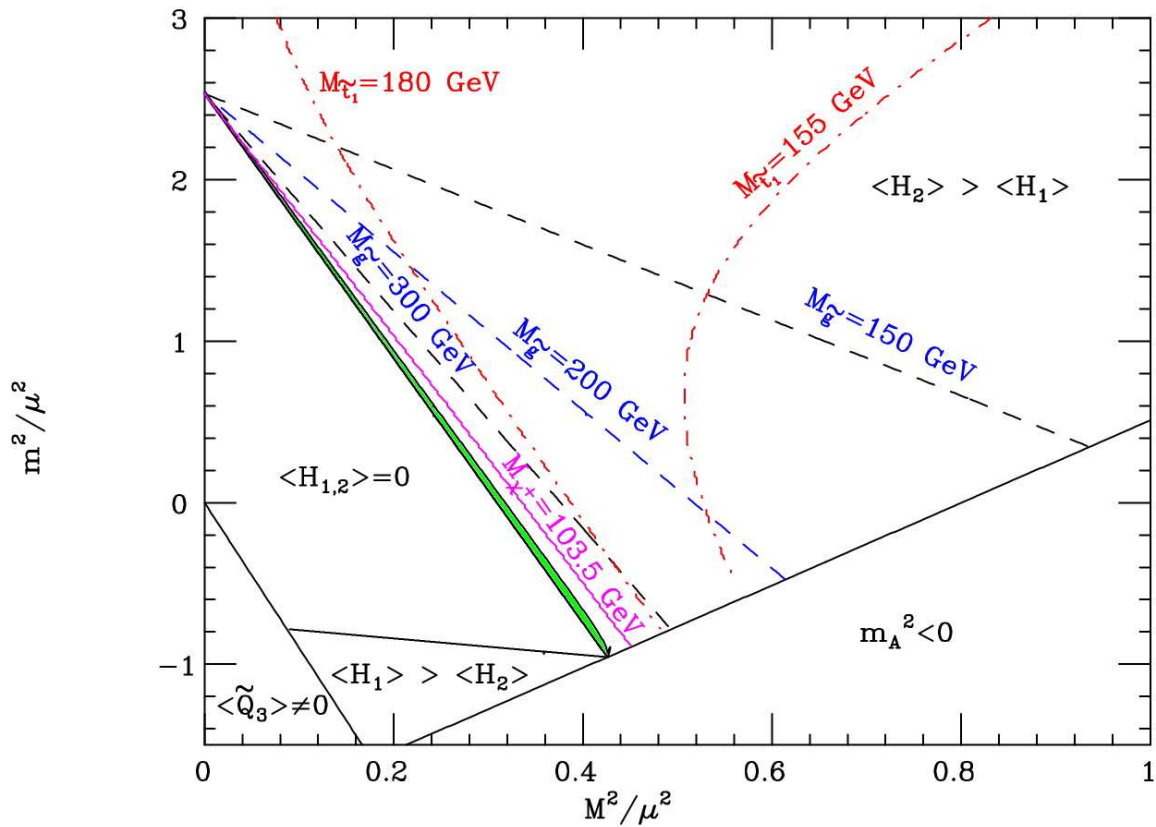}
\end{center}
\caption{ 
The phase diagram of the minimal supersymmetric SM, assuming a
universal scalar mass $m^2$, a gaugino unified mass $M$, a Higgsino mass
$\mu$, and trilinear term $A=0$, with all parameters defined at the GUT
scale. The top Yukawa coupling is fixed such that $m_t=172.7$~GeV and
$\tan\beta =10$ in the usual phase with electroweak breaking. Some
contours are shown for masses of the lightest stop ($M_{{\tilde t}_1}$),
the gluino ($M_{\tilde g}$), and the lightest chargino ($M_{\chi^+}$). The
green (gray) area shows the region of parameters allowed after LEP Higgs
searches.
}
\label{fig1}
\end{figure}

The phase diagram of the minimal supersymmetric SM is shown in 
\fig{fig1}. 
A first peculiarity of supersymmetry is the existence of phases where color and
electric charge are broken. This happens, for instance, at 
negative $m^2$ and small $M$, where the third-generation 
squark $\tilde Q_3$ gets a vacuum expectation value. Actually, assuming strict universality,
there is an even larger region where the selectron gets a vacuum expectation
value, which is not shown
in \fig{fig1} since, for the sake of argument, we take a common scalar mass only for the
particles involved in the conventional $SU(2)\times U(1)$ breaking pattern (third-generation
squarks and the two Higgses).
 
More interesting is a special multi-critical point, 
separating the various Higgs phases, that corresponds to 
vanishing Higgs bilinear terms ($m_1^2=m_2^2=m_3^2=
0$)\footnote{These three conditions cannot be in general satisfied in the case
of only two free parameters. However, \fig{fig1} corresponds to fixed $\tan\beta$,
and thus $m_3^2$ automatically vanishes, whenever $m_1^2=m_2^2=0$.}. 
This point, which is actually a surface in the case of general soft terms, occurs 
at negative $m^2$, in the example we are considering.
Moving away from the multi-critical point, different phases emerge, 
depending on the signs and the values of
$m_1^2$ and $m_2^2$ at the scale $M_S$. For positive $m_{1,2}^2$, the potential is
stabilized at the origin; the scale $M_S$ is larger than the critical
scale $Q_c$, and electroweak symmetry is unbroken. Notice that this phase (marked as
$\langle H_{1,2}\rangle =0$ in \fig{fig1}) extends, for $M=0$, to rather large values 
of $m^2/\mu^2$. This is a peculiarity of the assumption of strict universality which, together
with the known value of the top mass, leads to a certain cancellation of the contribution to
$m_Z^2$ proportional to $m^2$. Varying the top Yukawa coupling (or, ultimately, $\tan\beta$),
one can obtain higher degrees of cancellation, approaching what is known as ``focus 
point"~\cite{focus}. To compensate for this reduced dependence on $m^2$ (a characteristic
of universality, not shared by generic soft-term structures) we have expanded in \fig{fig1}
the scale of the vertical axis, with respect to the horizontal axis. For the same
reason, the precise location of the boundary between the broken and unbroken phases at small $M$
sensitively depends on the values of the coupling constants and on the degree of
accuracy of the calculation. In our figures, we have chosen $\alpha_s(m_Z)=
0.1176$, and fixed the top Yukawa coupling corresponding to
$m_t=172.7$~GeV and $\tan\beta =10$ in the broken phase. We have also limited
our RG evolution to one-loop approximation. 

In the limit of exact supersymmetry and PQ symmetry, all quadratic
Higgs terms in \eq{higgspot} vanish. Actually, since in supergravity scenarios
the PQ breaking can easily arise only from supersymmetry breaking~\cite{giumas},
we will refer to this case ($m_{1,2,3}^2=0$) as the supersymmetric limit. In this
limit, the Higgs potential has a flat direction $\langle H_1\rangle = \langle H_2\rangle$,
characteristic of supersymmetric $D$-terms. Supersymmetry breaking stabilizes this
direction as long as $m_1^2+m_2^2>2|m_3^2|$. If this is not the case, the Higgs field slides
up to the renormalization scale where the previous inequality is satisfied, as in
the Coleman-Weinberg mechanism. If $m^2/\mu^2
<-1$, this scale is actually larger than the GUT scale cutoff. At any rate, the important point
is that the Higgs vacuum expectation value is unrelated to the supersymmetry scale $M_S$ and in particular 
 $m_Z \ll M_S$. This region, which is of course experimentally ruled out
 is marked in \fig{fig1} by $m_A^2<0$.
Indeed, its boundary is characterized, in our analysis with fixed $\tan\beta$, by 
the condition $m_A\equiv m_1^2 +m_2^2 <0$
since, in the region of conventional electroweak breaking, $2m_3^2=\sin 2\beta (m_1^2
+m_2^2)$.

In the rest of the phase diagram in \fig{fig1} the Higgs vacuum expectation value is proportional to supersymmetry-breaking
terms, and therefore $M_S$ controls the size of electroweak breaking, thus providing a potentially realistic
solution to the hierarchy problem. Depending on whether $m_1^2$ or $m_2^2$ is driven negative, we
obtain two possible regions marked in \fig{fig1} as $\langle H_1\rangle > \langle H_2\rangle$
and  $\langle H_2\rangle > \langle H_1\rangle$, respectively. The first region, which occurs
only at negative $m^2$, has phenomenological difficulties in maintaining a perturbative 
top Yukawa coupling to large scales and in making the Higgs mass sufficiently heavy.
Therefore, we will concentrate on the region with $\langle H_2\rangle > \langle H_1\rangle$.

In this region, the inequality $Q_c>M_S$ is satisfied, and we can determine the overall 
mass scale of supersymmetric particles from the condition that radiative electroweak 
breaking reproduces the known value of $m_Z$. The complete mass spectrum can then
by computed at a given point of the phase diagram, and in \fig{fig1} we show some characteristic values of supersymmetric particle masses. 
In the bulk of the region, we find that supersymmetric colored particles weigh typically less
than 2--3 times $m_Z$, while some electroweak particles are lighter than $m_Z$. The values of
the supersymmetric masses have only mild variations in the bulk of the region, but they
precipitously increase in the proximity of the critical line separating the broken and unbroken
phases, where the critical scale $Q_c$ rapidly approaches $M_S$. Only near the boundary we can find
supersymmetric masses compatible with the present bounds from collider experiments. 
For instance, the chargino-mass LEP bound $M_{\chi^+}> 103.5$~GeV at 95 \% CL~\cite{lepchi}
rules out all the region to the right of the corresponding blue line in \fig{fig1}, allowing only the narrow strip 
between the blue and critical lines. Actually the negative Higgs searches impose even
stronger constraints on the allowed region. Taking into account the limits on Higgs 
production at LEP~\cite{lephs} in the channels $Zh$, $ZH$, $hA$, $HA$ (where $h$, $H$,
$A$ are the three neutral supersymmetric Higgses), we find that the only allowed points 
in \fig{fig1} are those inside the green (gray) region, clustering along the critical line.

A first conclusion that we can draw from these results is that the most natural prediction of
supersymmetry on the spectrum of new particles has already been ruled out, and only small
corners of parameter space are still allowed. This conclusion is of course well known and it
has been already quantified in different ways~\cite{natur,natur2}. Figure~\ref{fig1} presents an alternative way
to illustrate the problem.

However, \fig{fig1} also leads us to a new way of characterizing the allowed region, in terms
of criticality condition. The problem of understanding why supersymmetry may have chosen
highly untypical values of soft parameters, which appear to have the only effect to hide it
from collider searches, is now turned into the question of why supersymmetry wants to lie
in a near-critical condition. In the following, we will discuss possible statistical (or dynamical)
attempts to explain this puzzle. But before ending this section, we want to address the question of how general is our conclusion
that the only allowed parameter region of low-energy supersymmetry lies close to the critical
line, and we investigate if other regions, albeit tuned, can arise far from it.

\begin{figure}[t!]
\begin{center}
\includegraphics[width=16cm]{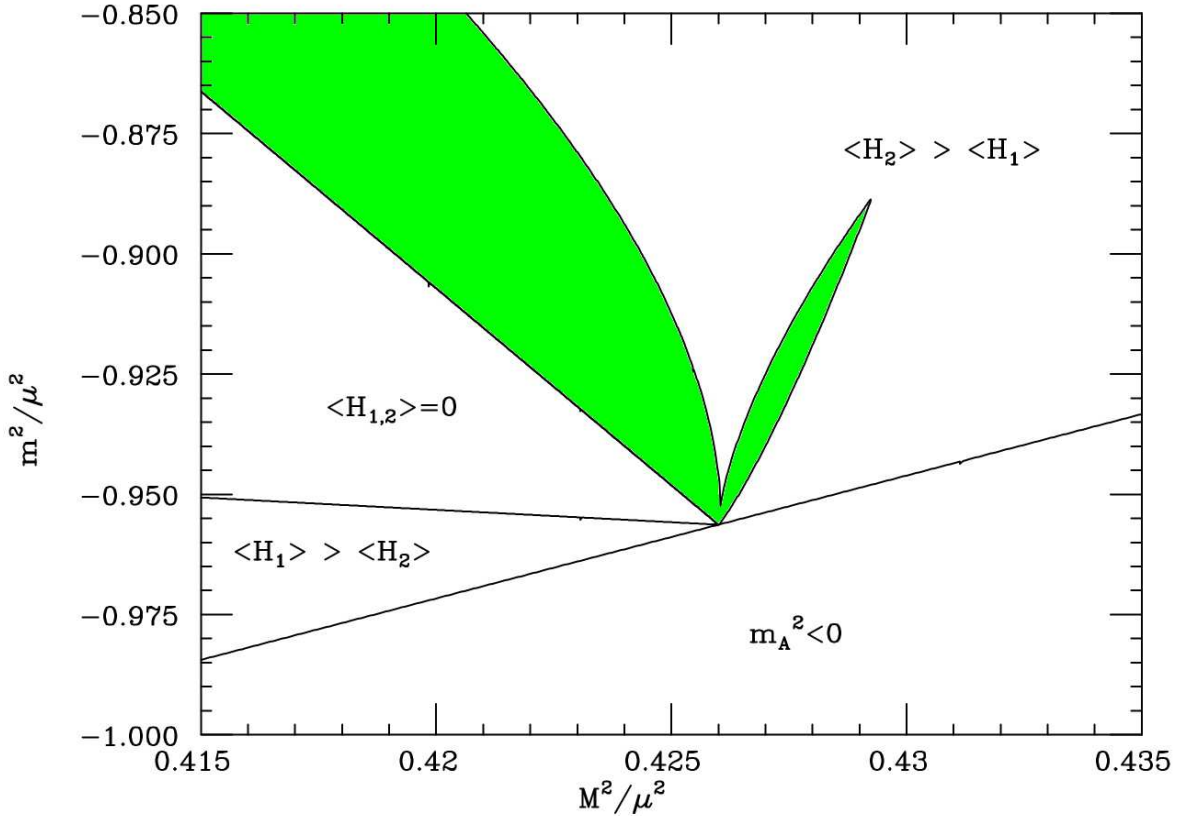}
\end{center}
\caption{ 
Same as \fig{fig1}, zooming in the allowed region where the Higgs pseudoscalar
mass is close to $m_Z$. 
 }
\label{fig2}
\end{figure}

It is well known that 
the experimental SM Higgs mass bound $m_h> 114.4$~GeV at 95\% CL~\cite{leph} 
does not directly apply to the 
supersymmetric case since, for a pseudoscalar mass $m_A$ near $m_Z$, the coupling of the
lightest Higgs boson to the $Z$ boson is reduced. In \fig{fig2} we zoom into the region of
parameter space where this happens. The thin sliver extending away from the multi-critical
point corresponds to the region allowed by LEP searches where $m_A\simeq m_Z$ and 
$m_h < 114$~GeV.
Besides the consideration that this region appears as a very special tuning of the underlying
parameters, we observe that it does not allow to depart significantly from the critical condition.
As explained in the appendix, the reason for this limited effect is that, to have a suppressed 
$hZZ$ coupling without conflict with the LEP searches in the 
$hA$ channel, one needs large corrections
to the Higgs quartic coupling. This requires again to be near criticality.

It is also known that the stringent lower limit on the stop mass derived from the Higgs-mass bound can be significantly relaxed for large values of the trilinear term $A_t$. This limit on $M_{{\tilde
t}_1} $ is more than 1 TeV for vanishing stop mixing, but it is reduced to 200--300~GeV when
the mixing reaches the condition $(A_t-\mu /\tan\beta )^2/(M_{{\tilde t}_1}M_{{\tilde
t}_2})=6$. This allows lighter stops and, apparently, less fine tuning. Indeed, if we increase
the value of $A$, the region allowed by Higgs searches becomes slightly larger than what
shown in \fig{fig1}. However, at the same time, large $A$ terms contribute to $m_Z^2$ and 
reduce the overall value of $M_S$. This has the effect of predicting a lighter supersymmetric
spectrum and push the mass contour lines of \fig{fig1} closer to the critical line. In this
case, the chargino mass limit plays the dominant role, and the allowed region is still clustered
along the critical line. 
A certain relaxation could be achieved if gaugino-mass unification does not hold, and if $M_2>M_3$
at the GUT scale. In this case, the chargino mass limit plays a more limited role, and we can
increase further the value of $A$ and make the Higgs boson heavier. However, this possibility
is limited by the bound on the stop mass.
In conclusion, we find that the connection between experimentally-allowed
supersymmetric parameters and criticality is robust under variations of the soft-term structure.

\section{Statistical Criticality}
\label{stat}
There have been various attempts to explain the tuning of low-energy supersymmetry 
by dynamical mechanisms or through extra symmetries~\cite{lith,other,other2,bs}. 
Ref. \cite{lith} marries the little Higgs idea to supersymmetry, suitably extending the
minimal model in order to make
one combination of the two Higgses a pseudo-Goldstone boson. The papers in ref. \cite{other},
by  providing extra contributions to the Higgs quartic coupling,
 focus just on the tuning produced by the $m_h$ bound.
The papers in refs. \cite{lith,other} represent departures from the minimal
supersymmetric SM right at the superparticle mass scale, which are
in principle testable at future colliders. However these models are rather complicated and it is hard to believe
that nature would choose such complication just to hide supersymmetry at LEP.
It is also fair to say that they do not fully solve the fine-tuning problem of supersymmetry. Notice in passing that this last
problem is also shared by the extension of the minimal model involving an extra Higgs
singlet superfield, unlike what is commonly stated but  as it has been recently 
emphasized in a detailed study \cite{Schuster:2005py}. In the models in ref. \cite{other2,bs} the theory retains the minimal field content
up to some ultra high scale, and the apparent tuning is supposedly explained by the  supersymmetry-breaking dynamics.
Ref. \cite{other2} represents a remarkable supergravity scenario where $Q_c$ is parametrically
tied to $M_S$, but it seems that the lifting potential, upon which this results is fully based, does
not have any sensible microscopic motivation~\cite{thaler}.
In sect.~\ref{dyn} we will 
 illustrate in more detail why the  dynamical explanation in ref. \cite{bs} has difficulties. 
 
 Here 
 we take a different approach and
try instead to provide  a  statistical explanation of criticality. We will be working under the multiverse
or landscape  hypothesis~\cite{lands}. According to this hypothesis, the fundamental 
description of nature features a tremendous multiplicity, 
a landscape, of physically inequivalent vacua and our local universe represents but one domain
of a multiverse.
% in  which the vacuum changes from domain to domain.
With the parameters of the low-energy effective field theory
changing from domain to domain, statistical considerations can be applied 
to deduce, under some assumptions, the likelihood of parameter
configurations. In particular,  observed properties of our domain, through their
physical relations to the parameters, some of which known and some of 
which unknown, can imply conditional probabilities on the unknown parameters.
Weinberg's prediction of the anthropically favoured size of the cosmological
 constant~\cite{Weinberg} is an example of that, with the existence of galaxies and the size
 of primordial density perturbations playing the role of the measured data of our domain.
 
 In analogy with Weinberg's approach to the cosmological constant problem, we will assume that
 the soft-supersymmetry breaking mass parameters are environmental quantities 
 varying across the multiverse.   Working in the context of hidden-sector 
 models, this means that each different vacuum in the landscape gives rise to a different 
 set of soft mass parameters $\{m_i\}$ up at a fixed scale, say at $M_P$.
 As the simplest  possibility, let first us assume that  
only the overall supersymmetric mass scale $M_S$ is
 environmental. More precisely let us assume that at the Planck scale the soft masses including
 $\mu$ are given by
 \be
 m_i = c_i M_S,
 \label{boundcond}
 \ee
 with the dimensionless coefficients $c_i$ fixed everywhere thoughout the landscape, while $M_S$ varies.
Let us also assume that  all the other dimensionless couplings 
(gauge and Yukawa) are fixed at the Planck scale.
It is possible to think of field theoretic   landscapes  that realize this condition, as we will
discuss in sect.~\ref{distrib}.  Let us consider the normal
situation in which the Higgs mass matrix is positive definite at the Planck scale. Under the above conditions
also $Q_c$ is fixed. Indeed the RG equations are homogeneous in the soft terms, so that  the RG evolution is written as  the evolution of the $c_i$ with $M_S$ constant. Then  $Q_c$, corresponding to the RG scale where 
\be
\det {\cal M}^2(Q)=m_1^2(Q)m_2^2(Q)-m_3^4(Q)\equiv M_S^4\left [c_1^2(Q)c_2^2(Q)-c_3^4(Q)\right ]
\ee
turns negative, depends on the high-energy scale $M_P$ and on the dimensionless couplings 
\be
Q_c= M_P \times F(c_i,\alpha_a)\, ,
\ee
but not on $M_S$. 
Here by $\alpha_a$ we collectively denote the gauge and Yukawa couplings at the Planck scale.
The physical values of the Higgs mass parameters are, in leading log approximation,
equal to the running masses computed at the RG scale $Q\equiv M_S$.
Two possibilities for the value of $M_S$ in the multiverse domain comprising our universe  are then given: {\it (i)}
$M_S> Q_c$, for which ${\cal M}^2$ is positive definite and thus $\langle H\rangle = 0$;
{\it (ii)} $M_S<Q_c$, for which ${\cal M}^2$ as at least one negative eigenvalue, implying $\langle H\rangle \not = 0$.

It is pretty clear we do not live in region {\it (i)}, and in fact it is not  even sure if in region {\it (i)}
there can exist anyone to ask this question \cite{anvev,nima}. Although a rich atomic structure may exist in this region \cite{anvev},
it  would look so different from our world that it seems rather unlikely it would be hospitable to life.
Moreover, and more simply, it has been shown \cite{nima} that
for $\vev{H}=0$ any primordial baryon density is very efficiently converted into leptons (mostly neutrinos) 
by electroweak sphalerons. These effects are now active down to temperatures of the order of $\Lambda_{\rm QCD}$,
at which conversion of baryons into leptons is energetically favored.
This feature of the $\vev{H}=0$ universe seems rather solid as it does not depend very much on the Yukawa
couplings of quarks and leptons (as long as they remain weak).  Therefore region 
{\it (ii)} is also strongly favored over region {\it (i)} 
for anthropic reasons.

Compatibly with the prior that we must live in region {\it (ii)}, 
we can ask what is the most likely value we expect $M_S$ to have.
The problem is phrased in complete analogy
with Weinberg's approach to the cosmological constant, with $\langle H\rangle \not = 0$ replacing the datum that galaxies exist. Then, 
under the assumption that the distribution
of $M_{S}$ is reasonably flat and featureless, and 
not peaked at $M_S=0$, we expect $M_S \sim Q_{c}$. For instance, as we will show in 
sect.~\ref{distrib},
in simple field theoretic
modelling of the landscape, a typical expectation is that the number of vacua 
with $M_S<m$ grows like a positive power $m^n$. Then, treating all vacua as equally 
probable, the prior $\langle H\rangle \not = 0$ leads to a conditional probability
\beq
dP= \left\{ \begin{array}{cl}
n\left (\frac {M_S}{Q_c}\right )^n \frac{dM_S}{M_S} & ~~~{\rm for}~~M_S<Q_c\\
0 & ~~~{\rm for}~~M_S>Q_c \end{array} \right . .
\label{prob}
\eeq
By this equation the average logarithmic separation of scales
is given by $\langle \ln Q_c/M_S\rangle = 1/n$, which agrees with the rough expectation 
$M_S \sim Q_c$  for a smooth distribution with $n= O(1)$. As the RG evolution {\it slowly} proceeds by 1-loop effects, the Higgs mass matrix ${\cal M}^2$ typically develops
 a {\it small} $O(\alpha)$ negative eigenvalue at the scale  $M_S\sim Q_c$, where the 
running is frozen. The weak scale will correspondingly be parametrically smaller than $M_S$. 
By minimizing the scalar potential at leading order in $L_S=\ln Q_c/M_S$, we have 
 \be
 m_Z^2 (\cos^22\beta +\delta \sin^4\beta)=\left .\frac{2}{m_1^2+m_2^2}\frac{d(\det {\cal M}^2)}{d \ln Q}\right |_{Q=Q_c} L_S+O\left( \frac{\alpha^2}{16\pi^2} L_S^2\right) ,
 \label{derivative}
 \ee 
% \be
% m_Z^2 (\cos^22\beta +\delta \sin^4\beta)=-2\left .\frac{\frac{d(\det {\cal M}^2)}{d \ln Q}}{m_1^2+m_2^2}\right |_{Q=Q_c} L_S+O(\frac{\alpha^2}{16\pi^2} L_S^2)
% \ee 
%\be
% m_Z^2 (\cos^22\beta +\delta \sin^4\beta)=-2\left .\frac{{d(\det {\cal M}^2)}/{d \ln Q}}{m_1^2+m_2^2}\right |_{Q=Q_c} L_S+O(\frac{\alpha^2}{16\pi^2} L_S^2)
% \ee 
where $\delta$, defined in the appendix, represents the top-stop quantum correction to the Higgs quartic coupling and
where $\sin 2 \beta=2m_3^2/(m_1^2+ m_2^2)$.
For $\tan\beta\gsim 5 -10$, the above equation is well approximated by its $\tan\beta\to \infty$ limit
\bea
m_Z^2(1+\delta)&\simeq &2\frac {d m_2^2}{d\ln Q} L_S \\
&=& \left [\lambda_t^2\left (m_{\tilde t_L}^2+m_{\tilde t_R}^2+|A_t|^2\right )
-\frac{g_1^2}{5} \left (M_1^2+\mu^2\right )
 -g_2^2 \left (M_2^2+\mu^2\right )\right ] \frac{3L_S}{4\pi^2}\, .
 \label{littlehie}
\eea
By assuming the stop parameters to dominate the above
equation with $m_{\tilde t_L}^2\sim m_{\tilde t_R}^2\sim |A_t|^2\sim M_S^2 $ and 
by using $\langle \ln Q_c/M_S\rangle = 1/n$, we find the following average little hierarchy
\be
\left\langle \frac {m_Z^2}{M_S^2}\right\rangle
\,\simeq\, \frac{9\lambda_t^2}{4\pi^2(1+\delta)}
\,\times \frac{1}{n}\,\simeq 0.15\left( \frac{114\gev}{m_h}\right)^2 
\,\times \frac{1}{n}.
\label{ratio}
\ee
Notice that while the loop factor in \eq{ratio}
helps to explain the
little hierarchy problem in supersymmetry, it 
still falls short to explain it completely. Indeed, the Higgs mass limit
requires stop masses close to 1~TeV, a therefore a little hierarchy
$m_Z^2/M_S^2 \lsim 0.01$--$0.1$. However,
 as we will
argue in sect.~\ref{distrib}, 
it is rather reasonable to imagine field theoretic landscapes  where $n$ is somewhat bigger than 1, say $O({\rm a \,few})$, though not much
bigger. For instance if there are $O(10^{500})$ vacua, 
as perhaps suggested by string theory~\cite{vacua}, and if $M_S$ can range up to 
$M_P$, then $n< 30$. So it is reasonable
for the ratio in eq.~(\ref{ratio}) to be between $0.01$ and $0.1$ but not much smaller,
thus providing an argument  why supersymmetry should be elusive at LEP but not at the LHC.
Of course there has been a price to pay. Supersymmetry  looks tuned because throughout the landscape it is  much more likely  to be in the region with  $\langle H \rangle  = 0$ 
 than in the region  $\langle H \rangle\not = 0$ : the most likely points with
 $\langle H \rangle \not = 0$ are then  close to the boundary of the two regions, where a little hierarchy is present.
 
We should stress that these conclusions are based on statistical arguments
and that the average in \eq{ratio} has a variance of the same order,
{\it i.e.} $(\vev{X^2}-{\vev{X}}^2)/{\vev{X}}^2=1$ with $X\equiv m_Z^2/M_S^2$.
 In sect.~\ref{pheno} we will discuss in a little more detail 
the natural ranges of particle masses
in this scenario.
For the remaining part of this section we want instead
 to address some technical and conceptual questions.
 
 Let us address a technical question first.
 We have so far been working with 1-loop RG evolution. Accordingly we should be controlling
 only leading log effects $\sim (\alpha \ln)^n$ and neglect finite threshold 
effects. On the other hand, we have used as part of our logic the RG evolution 
between two scales $Q_c$ and $M_S$ that practically coincide, $\ln Q_c/M_S 
\sim 1/n\lsim 1$, in apparent contradiction to our leading-log approximation.
Our results are nonetheless correct. Consider indeed the physical supersymmetric 
mass parameters
as computed using RG evolution and thresholds to all orders (our boundary conditions at $M_P$ should
also be given within some renormalization scheme)
\be
m_i^2|_{\rm phys}= M_S^2 F_i(M_S/M_P,\alpha_a,c_j) .
\label{exact}
\ee
 Now we can define $Q_c$ as the critical value of $M_S$ below which the Higgs mass matrix develops a 
 negative eigenvalue. With this definition we can go back and follow the same logic from above eq.~(\ref{prob}). By varying $M_S$ the $F_i$ vary, at leading order in $\alpha_a$,  according to the 1-loop
 RG equation and all our results follow, to  that accuracy. Notice that
 $Q_c$ shifts by $O(1)$ when going from leading to next-to-leading order, 
but by the above definition $\ln Q_c/M_S$ is a  scheme-independent physical quantity.
%  The point is that,
%   even though  next-to-leading order effects determine $O(1)$ shifts in $\ln Q_c$, the relative change of the first derivative in eq.~(\ref{derivative})  is only $O(\alpha)$. 
%   the logarithm $\ln M_S/Q_c$ has a robust physical meaning.
 
Another issue concerns the dependence of our conclusions on the ``choice" of priors. We have used a  weak prior corresponding to
the rather weak request that $SU(2)_L\times U(1)$ be dominantly broken by an elementary scalar field.
In anthropic considerations
the interesting  implications of a stronger prior, normally called
the Atomic Principle, have also been studied~\cite{anvev}. The Atomic Principle
is based on the remark that the existence of a rich 
spectrum of nuclei and atoms, crucial for the existence of the complex world we live in, severely constrains the ratio of the electroweak and strong scales $v_F/\Lambda_{\rm QCD}$. In particular it is rather safe to argue that
if $v_F$, keeping every other coupling fixed, were just a factor 5 bigger than its actual value, then 
neutrons decay inside nuclei and no complex chemistry is possible.
For $v_F$ significatively smaller than its actual value, it is harder to 
control exactly what happens but, as we already said above, atomic physics is drastically modified, and moreover for $v_F\lsim
\Lambda_{\rm QCD}$ it seems very difficult to have enough baryons \cite{nima}. While this last
bound seems rather robust even when  the other parameters are varied,  the upper bound on $v_F$
can in principle be lifted by allowing suitable variations of both Yukawa couplings and cosmological parameters.
For instance it has been recently pointed
out \cite{Harnik:2006vj} that the upper bound on $v_F$ can be fully relaxed, thus allowing a  so-called  Weakless Universe, once some cosmological parameters are modified.
As we shall see in a  moment, under very reasonable assumptions, our
results do not crucially depend on the upper bound on $v_F$, so that 
the existence or the absence of this bound do not really affect our conclusions.
 In what follows  we will simply indicate by Atomic Principle  the request that $v_F\lsim 10^3\Lambda_{\rm QCD}$ while by
Weak Principle we will indicate the basic request $\langle H\rangle \not =0$.

If the Weak Principle is taken, like we did so far, as the
only relevant prior, then the value of the $Z$ mass in our patch is expected to be typical of the set of patches where the electroweak symmetry is broken, and therefore $\vev{m_Z}\sim 90$~GeV.
Since we have found that, up to the little-hierarchy factor, $Q_c\sim \vev{m_Z}$, and since 
 $\Lambda_{\rm QCD}$ is fixed to $1 \GeV$ throughout the multiverse, we conclude that
the presence of atoms is a fundamental and 
generic feature of practically all  vacua  breaking electroweak symmetry. Parametrically this corresponds to the fact that the fundamental ratio $Q_c/\Lambda_{\rm QCD}$ does not scan and it is fixed to
the right value in our theory.
However, if also the Atomic Principle is taken as a prior, then, by definition, our local patch is not expected to be typical among those where electroweak symmetry is broken
 and the critical scale $Q_c$ will  no longer be around the TeV.

Consider the case $Q_c\gg 1$ TeV and
write \eq{derivative} as  $m_Z^2=\alpha M_S^2\ln (Q_c/M_S)$, where 
$\alpha$ describes a one-loop factor depending on coupling constants and dimensionless
soft-term ratios. The  Weak and Atomic  Principles correspond to the condition
$0<m_Z<10^3\Lambda_{\rm QCD}\equiv \Lambda_W$, and 
therefore restrict the distribution of vacua 
\beq
dN\propto
dM_S^n 
\eeq
only to values of $M_S$ in the two domains
\beq
{\cal D}_1:~0<M_S<x_1 Q_c,~~~
{\cal D}_2:~x_2Q_c<M_S<Q_c .
\label{prob2}
\eeq
Here $x_{1,2}$ are the solutions of $\alpha x^2\ln x=-\Lambda_W^2/Q_c^2$ with $x_1<x_2$.
We are considering
the case of small $\Lambda_W^2/Q_c^2$, where $x_1=O(\Lambda_W /Q_c)$ and $x_2$ is close to
unity, $x_2\simeq 1-\Lambda_W^2/(\alpha Q_c^2)$. 

The relative number of vacua in the two domains $ {\cal D}_1$ and ${\cal D}_2$ is
\beq
\frac{\int_{{\cal D}_1} dN}{\int_{{\cal D}_2} dN}=\frac{x_1^n}{1-x_2^n}
\simeq \frac{\alpha}{n}\left( \frac{\Lambda_W}{Q_c}\right)^{n-2}.
\eeq
Therefore, for $n<2$ vacua in ${\cal D}_1$ dominate, while the region ${\cal D}_2$
is favored for $n>2$. The average value of the ratio between the weak and supersymmetric
scales in region ${\cal D}_2$ (for $n>2$) is
\beq
\left\langle \frac{m_Z^2}{M_S^2}\right\rangle_{{\cal D}_2}\equiv
\frac{\int_{{\cal D}_2} dN ~\alpha \ln Q_c/M_S}{\int_{{\cal D}_1\cup{\cal D}_2} dN}
\simeq \frac{\Lambda_W^2}{2Q_c^2}.
\eeq
In these vacua there is a huge hierarchy between $m_Z$ and $M_S$, and the 
low energy theory is either the SM or Split Supersymmetry~\cite{split}, depending on the masses of  
higgsinos and gauginos. In region ${\cal D}_1$ (for $n<2$), we find
\beq
\left\langle \frac{m_Z^2}{M_S^2}\right\rangle_{{\cal D}_1}\simeq \alpha
\ln \frac{Q_c}{\Lambda_W},
\eeq
and the low-energy theory is just ordinary untuned supersymmetry. We never encounter the
situation where a little hierarchy $m_Z^2/M_S^2\sim \alpha$ is favored, unless
we artificially tune $\Lambda_W\simeq Q_c$.

%Consider the case $Q_c\gg 1$ TeV.  There will be two regions of $M_S$ satisfying the Atomic and Weak principles.
%One is for $M_S$ very very close to $Q_c$, namely
%\be
%Q_c^2\left [1 -\left (\frac{\TeV}{\alpha Q_c}\right )^2\right ] < M_S^2< Q_c^2
%\ee
%where we have sloppily written the AP as $m_Z^2=\alpha M_S^2\ln (Q_c/M_S)< (\TeV)^2$. In these vacua there is a huge hierarchy between the weak scale and $M_S$, and the low energy theory is either the SM or Split Susy, depending on the masses of  higgsinos and gauginos.
% The number of vacua of this type scales like $N_{Split}\sim Q_c^{n-2}(\TeV)^2/\alpha$. The other region is for 
% $M_S\lsim (\TeV)\ll Q_c$, where the low energy theory is just ordinary untuned supersymmetry. The number of vacua 
% of this second type scales like $(\TeV)^n$. For $n>2$  the Split type vacua dominate while for $n<2$ we have that ordinary untuned supersymmetry is favored. In no way is the situation $m_Z^2/M_S^2\sim \alpha$ favored.
 
 The strong dependence of our conclusions of the number of  priors is not surprising, 
 given that we  only have one aleatory variable $M_S$ at hand. So in order to reach a more robust conclusion we should 
 consider the general situation where also some other parameter is scanned.  The 
Atomic Principle depends crucially
 on $\Lambda_{\rm QCD}$ and so we will assume this parameter to be scanned in some 
range. The scanning of $\Lambda_{\rm QCD}$ is given by a corresponding 
scanning of $\alpha_s$ up at the Planck scale. Since $\ln \Lambda_{\rm QCD}/M_P
\propto -1/\alpha_s$ and since it is reasonable to expect $1/\alpha_s$ to be scanned roughly linearly, we will assume the measure to be $\propto d\ln \Lambda_{\rm QCD}$.  Of course it is hard to imagine $Q_c$ not to scan when $\Lambda_{\rm QCD}$ does.
Actually, as also $Q_c$ is related to dimensional transmutation, we will assume a measure
for the distribution of vacua
 \be
 d N\propto  dM_S^n \, d\ln Q_c\, d\ln \Lambda
 \label{metr}
 \ee
 where, for convenience, we defined $\Lambda\equiv \Lambda_W/\sqrt{\alpha}$.
The Weak and Atomic Principles restrict the acceptable vacua in the region
\beq
{\cal D}: ~0<M_S<Q_c,~Q_c<\bar Q_c,~M_S\sqrt{\ln Q_c /M_S}<\Lambda <\bar \Lambda,
\eeq
where $\bar Q_c$ and $\bar \Lambda$ are the maximum values of $Q_c$ and $\Lambda$ over the landscape. We take $\bar \Lambda \gsim \bar Q_c$, to allow a sufficient scanning range for
$\Lambda_W$.

Since we are mainly interested in the distribution of $r\equiv M_S/Q_c$, it is convenient to
integrate \eq{metr} over the other variables in the
 region $\cal D$ admitted by the Weak and Atomic Principles. We find
 \be
 dN \propto \frac{\bar Q_c^n}{n}\left [\ln \left (\frac{\bar \Lambda}{\bar Q_c r}\right )-\frac{1}{2}\ln\ln\frac{1}{r}+\frac{1}{n}\right ]dr^n .
 \ee
Aside from a
 mild $\ln (1/r)$ dependence, the above measure describes a probability function essentially
 identical to \eq{prob}. Indeed, we find an average little hierarchy
 \beq
\left\langle \frac{m_Z^2}{M_S^2}\right\rangle \equiv
\frac{\int_{\cal D} dN ~\alpha \ln Q_c/M_S}{\int_{\cal D} dN}=\frac{\alpha C}{n},~~~~~
C=1+\frac{2-n}{4+n\left( \gamma +\ln n \bar\Lambda^2/\bar Q_c^2 \right)},
\eeq
 where $\gamma$ is the Euler constant. Therefore, up to an $O(1)$-coefficient $C$, we obtain
 the same result as in \eq{ratio}, corresponding to the case  
 where only $M_S$ scans and where only the Weak Principle is used.
 In practice by integrating over a logarithmically distributed $\Lambda_{\rm QCD}$ we have ``integrated out" the Atomic
 Principle. Finally notice that scanning $\Lambda_{\rm QCD}$ is crucial. If we scanned only $Q_c$, we would not reach this conclusion. In that case the distribution of $r$ would end up being
 \be
 dN \propto \frac{d\ln\frac{1}{r}}{(\ln\frac{1}{r})^{n/2}}
 \ee
 showing once again that, depending on $n$ large or smaller than $2$, either Split or untuned 
 supersymmetry are favored.
 
 This discussion also partially illustrates  
 the result that will be obtained with an independent scan of all soft terms. If the scan is restricted to a parameter region such that the
 squared masses for all scalars are positive at the high-energy scale, then $Q_c$ has a maximum, which
 acts as an upper bound on $M_S$ under the Weak Principle. Again, our considerations will
 lead to a little hierarchy in $m_Z^2/M_S^2$. The case of an independent scan of the higgsino mass
 parameter $\mu$ is particularly interesting, and it will be the subject of sect.~\ref{scamu}.
 
 Our results so far have been based on the standard soft term scenario, where the Higgs squared mass
 matrix starts out positive at the high-energy scale and develops a negative eigenvalue while flowing to lower energy scales.
 In this scenario we have argued that, when large values of $M_S$ are statistically favored, then the Weak Principle
 implies   criticality of the physical Higgs mass. Large values of $M_S$ are generically  favored
 when supersymmetry is broken at tree level, as discussed in sect.~\ref{distrib}.  
 On the other hand,  dynamical  supersymmetry breaking is an interesting possibility which suggest that $M_S$ could be
 logarithmically distributed or even peaked at low values. It is easy to see that also in this second case, where
 \beq
 dN = d(1/M_S^n)
 \label{1on}
 \eeq
 with $n>0$, the Weak Principle can lead to criticality. However, this will require
 the remarkably  unusual situation where ${\cal M}^2$ has a negative eigenvalue in the 
 ultraviolet and flows to become positive definite
 in the infrared. We can again define as $Q_c$ the RG scale where this happens. Then the  sign of $\ln {Q_c/M_S}$ and of $d m_2^2/d\ln Q$ in eq.~(\ref{littlehie}) will both be reversed with respect to the previous case. The Weak Principle and eq.~(\ref{1on})
 imply $\langle \ln Q_c/M_S\rangle = -1/n$ meaning that RG evolution gets frozen just before ${\cal M}^2$ turns positive.
The phenomenology of this scenario has been recently discussed in 
ref.~\cite{nega}.

  \section{Scanning $\mu$}
\label{scamu}
Among the soft parameters the role of $\mu$ is special, as it does not break supersymmetry
while it breaks a global PQ-symmetry which is respected by the other soft parameters.
This properties of $\mu$  make its origin often problematic from the model building viewpoint.
This is generically the case in models with gauge or anomaly mediated supersymmetry breaking.
On the other hand, in ordinary tree-level gravity mediation, the size of $\mu$ (and of $m_3^2\equiv B\mu$) can be naturally associated to $M_S$~\cite{giumas}. In this section we will  
present an alternative viewpoint on the $\mu$ problem, proposing a solution based on 
statistical considerations, obtained by exploring the implications of
scanning $\mu$ independently of $M_S$ over the landscape. This will on one side illustrate
some features of the general case in which all soft terms are independently scanned, but on the other side
it will remarkably predict a favored range for the size of $\mu$ and of $\tan\beta$.

We consider a slight modification of our previous ansatz at the Planck scale with
\be
\mu=\mu_0\qquad\qquad m_3^2=b M_S\mu_0 
\ee
and  all other soft terms given by eq.~(\ref{boundcond}), taking $\mu_0$ and $M_S$ to
scan independently, while  $b$ is fixed.
Writing the running Higgs mass matrix as
\be
{\cal M}^2=\left (\begin{array}{cc}
{\tilde m}_1^2+\mu^2& B\mu\\
B^*\mu& {\tilde m}_2^2+\mu^2
\end{array}\right ) ,
\ee
the condition of criticality is
\be
\det {\cal M}^2={\tilde m}_1^2{\tilde m}_2^2 +({\tilde m}_1^2+{\tilde m}_2^2-|B|^2)\mu^2+\mu^4=0\, .
\label{crlimu}
\ee

\begin{figure}[t!]
\begin{center}
\includegraphics[width=16cm]{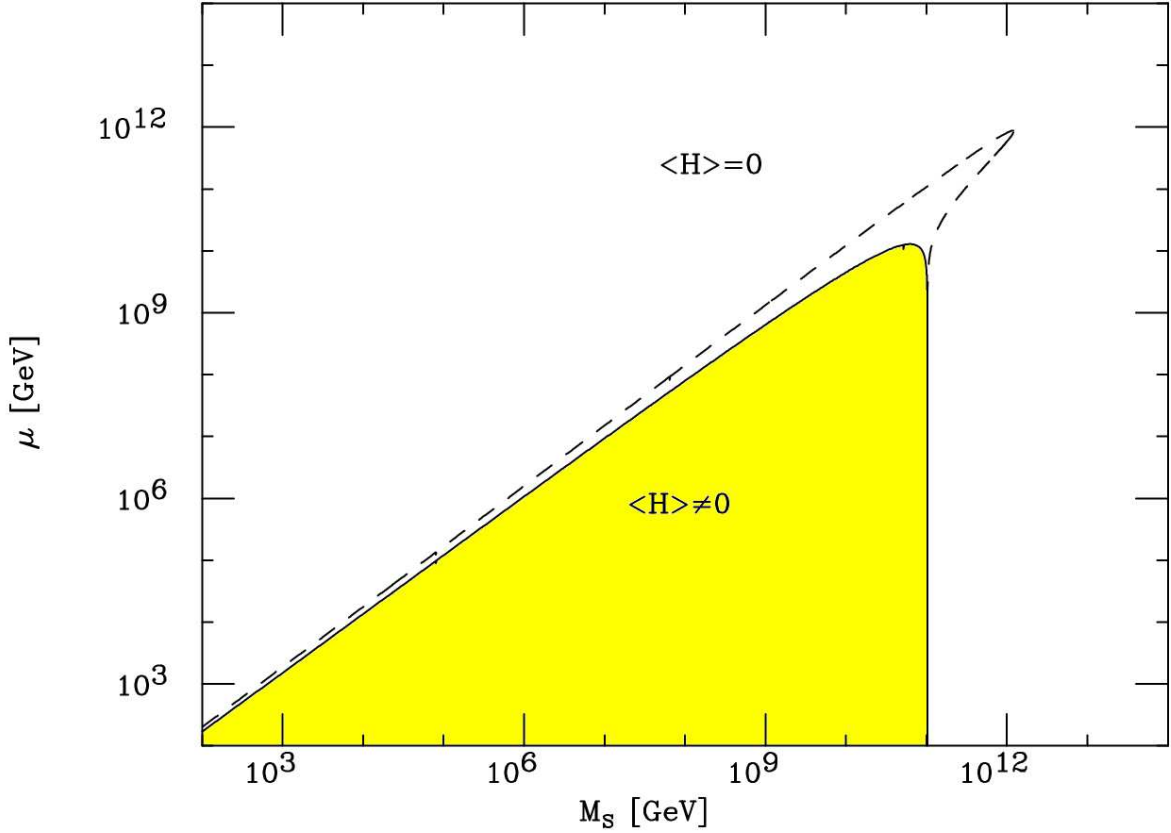}
\end{center}
\caption{ The critical line separating the broken and unbroken phases. Here $\mu$ is
defined at the scale $M_S$ and we have fixed the top Yukawa coupling corresponding to
$m_t=172.7$~GeV for large $\tan\beta$. Scalar universality and gaugino unification is 
assumed with boundary conditions at the GUT scale $m^2=M^2=M_S^2$, $A=0$ and
$B=0$ (solid line), $B=\sqrt{2}M_S$ (dashed line).
}
\label{fig4}
\end{figure}

The critical line in the $\mu$--$M_S$ plane, corresponding to \eq{crlimu}, is shown in \fig{fig4}, 
in the case of scalar universality and gaugino unification with $m=M=M_S$, $A=0$ and
$B=0$ (solid line), $B=\sqrt{2}M_S$ (dashed line) at the GUT scale. To understand the
shape of these curves,
let us indicate by $\bar Q_c$ the critical RG scale for $\mu=0$ and let us 
start with the case in which
${\tilde m}_1^2+{\tilde m}_2^2-|B|^2>0$ for $Q>\bar Q_c$ (as for the solid line of \fig{fig4}). 
Then, in ordinary RG flows with $d {\tilde m}_2^2/d\ln Q>0$,
an increase in $\mu$ will lower the value of $Q_c$. Let us study the critical line close to $M_s=\bar Q_c$. 
The small eigenvalue of ${\cal M}^2$ is 
\be
\lambda_{small}\simeq {\tilde m}_2^2+(1-r_B)\mu^2+O(\mu^4) ,
\ee
where $r_B=|B|^2/{\tilde m}_1^2$. By using the qualitative behaviour $2{\tilde m}_2^2=
\alpha M_S^2\ln M_S/\bar Q_c$, 
the critical curve is then given by
\be
\lambda_{small}=0\quad\Rightarrow\quad \mu^2\simeq \frac{\alpha}{2(1-r_B)} M_S^2\ln \bar Q_c/M_S \equiv f^2(M_S) .
\label{mucurve}
\ee
This equation implies that in the region $M_S\sim \bar Q_c$, which is favored when the number of vacua grows with $M_S$, there exist a moderate $O(\alpha)$ hierarchy also between $\mu^2$ and $M_S^2$. To see how this works 
explicitly, let us make the simple assumption on the distribution of vacua
\be
dN\propto dM_S^n d\mu^m
\ee
with $m,n>0$. The Weak Principle restricts the acceptable vacua to lie in the region
\beq
{\cal D}_\mu :~ 0<M_S<\bar Q_c,~0<\mu<f(M_S).
\eeq
Since $m_Z^2\simeq -2\lambda_{small}$, the average
ratio $m_Z^2/M_S^2$ is given by
 \beq
\left\langle \frac{m_Z^2}{M_S^2}\right\rangle \simeq
\frac{\int_{{\cal D}_\mu} dN ~\left[ \alpha \ln \frac{Q_c}{M_S}-2(1-r_B)\frac{\mu^2}{M_S^2}\right]
}{\int_{{\cal D}_\mu} dN}=\frac{\alpha}{n+m}.
\eeq
This shows that  an independent scanning of $\mu$ changes \eq{ratio} simply  by 
the replacement $n\to n+m$. The extra suppression is easy
to understand, since a positive $m$ pushes $\mu$ towards the critical line, as a positive $n$
pushes $M_S$. The average ratio $\mu^2/M_S^2$
is instead
\be
\left\langle \frac{\mu^2}{M_S^2}\right\rangle = \alpha ~\frac{m}{4(1-r_B)(n+m)}.
\ee
This shows that higgsinos are lighter than the typical supersymmetric particles by the square
root of a loop factor.

Using $\sin 2\beta=2B\mu /(m_1^2+m_2^2)$, we can express the average value of $\tan\beta$ as
\beq
\left\langle \tan\beta \right\rangle \simeq \frac{1}{\sqrt{r_B}}\left\langle \frac{M_S}{\mu} \right\rangle .
\eeq 
Therefore,   $\tan\beta\sim 1/{\sqrt \alpha}\sim 5$--$10$ is the most natural expectation in this framework. This
is welcome, because $|\cos 2\beta |\simeq 1$ plays a non-negligible role in pushing the mass of the lightest Higgs above the LEP bound.
The generic prediction with soft parameters of the same order of magnitude
is $\tan\beta = O(1)$, and it is well known that large $\tan\beta$ can be obtained only with
a fine tuning of order $1/\tan\beta$ in $B\mu$.  In our scenario, $\tan\beta\gg 1$ is just
an added bonus of statistical criticality.

So far we have considered the case in which ${\tilde m}_1^2+{\tilde m}_2^2-|B|^2$ is always
positive for $Q>\bar Q_c$. If this is not the case, the coefficient of $\mu^2$ in \eq{crlimu} becomes negative before ${\tilde m}_2^2$,
in the RG evolution starting from $M_{GUT}$. Therefore we expect that the
most probable values of $\mu/M_S$ are of order unity, with no loop-factor suppression. This is
confirmed by the result shown by the dashed line in \fig{fig4}, which corresponds to the case of a
large $B$.

In conclusion, the Weak Principle and statistical considerations based on an independent  
scan of $M_S$ and $\mu$ 
offer a solution to the $\mu$ problem, since the most
probable values of $\mu$ turn out to be close to $M_S$. Moreover, in the case of positive
${\tilde m}_1^2+{\tilde m}_2^2-|B|^2$ for $Q>\bar Q_c$ (which is actually the most likely situation for
typical soft terms of comparable sizes), our anthropic assumption gives the testable prediction
that both $\mu/M_S$ and $1/\tan\beta$ are of the order of the square of a loop factor. 

 \section{Dynamical Criticality}
 \label{dyn}
 Our environmental argument to explain (if not  post-dict) the little hierarchy 
 of the minimal supersymmetric SM 
 followed very closely Weinberg's approach to the cosmological constant problem. 
 Like for the cosmological constant,
we think it is instructive to see why a dynamical mechanism, where 
$M_S$ is a dynamical rather than aleatory variable, is difficult to be realized.

  It is well known that directions which are flat in the supersymmetric limit can
 dynamically generate 1-loop hierarchies via the Coleman-Weinberg mechanism. Indeed,
 in the presence of soft supersymmetry breaking, the effective potential along a flat direction
 $\phi$ can be written in general as $V= m^2(\phi)\phi^2$, where $m^2$ is a running effective
 mass squared. If, starting from a positive value at some high-energy scale, $m^2$ crosses 
zero at some
smaller   $\phi_c$, then $m^2(\phi )=\alpha M_S^2 \ln \phi /\phi_c$ and
the minimum of the potential will be at $\phi =\phi_c/\sqrt{e} $, where $e$ is Napier's 
constant. At this minimum, $m^2$ is one-loop suppressed with respect to the typical soft mass scale
thus dynamically realizing a little hierarchy. Barbieri and Strumia \cite{bs} have proposed a set up
where a Coleman-Weinberg potential explains the little hierarchy. Following ref.~\cite{noscale} (see
also ref.~\cite{slop}),
these authors
have considered the situation in which the overall supersymmetric scale $M_S$ in eq.~(\ref{boundcond}) is itself a scalar field, a modulus, with respect to which the potential should be minimized. The latter can generally be written as
\be
V(M_S,H_1,H_2)=V_{MSSM}(M_S,H_1,H_2)+V_S(M_S) ,
\ee
where $V_{MSSM}$ represents the ordinary potential of the supersymmetric SM, with terms quadratic and quartic in the Higgs fields and with the running soft mass matrix ${\cal M}^2$. If, for some reason, $V_S$ could be neglected, then the minimization
of $V_{MSSM}$ would dynamically realize an $O(\alpha)$ hierarchy between $\vev{H}^2$
and $M_S^2$. Indeed, under the same assumptions
of sect.~\ref{stat}, $V_{MSSM}$ is strictly positive for $M_S>Q_c$. For $M_S<Q_c$
we can minimize $V_{MSSM}$ with respect to $H_{1,2}$ thus yelding an effective
negative definite potential for $M_S$
\be
V_{MSSM}^{\rm eff}(M_S)= -\frac{2M_S^4}{g_2^2+g_1^2}~ \frac{\left[
c^2_1(M_S)c^2_2(M_S)-c_3^4(M_S)\right]^2}{\left[ c^2_1(M_S)+c^2_2(M_S)\right]^2-4c_3^4(M_S)},
\ee
where, as before, the Higgs mass parameters are expressed as $m_i^2=c_i^2M_S^2$ $(i=1,2,3)$.
Here to simplify the notation we have neglected the top-stop corrections to the Higgs quartic
coupling. Expanding the $c_i$ in $\ln Q_c/M_S$, we find (indicating schematically the powers of loop factors
for later use)
\be
V_{MSSM}^{\rm eff}(M_S)= -\frac{\alpha}{(4\pi)^3}M_S^4\ln^2\frac{Q_c}{M_S}
\left[1+O\left( \frac{\alpha}{4\pi} \ln \frac{Q_c}{M_S}\right) \right] ,
\label{effective}
\ee
which, at leading order in $\alpha$, is minimized at $\ln Q_c/M_S=1/2$. Compared to our eq.~(\ref{ratio}), this result correspond to dynamically predicting $n=2$ which, in principle, is a
testable relation among soft parameters at the weak scale. It falls short, as already explained,
to fully account for the little hierarchy, but nonetheless it is a remarkable result. 

Unfortunately this result totally rests on our assumption of negligible $V_S(M_S)$. Now, $V_S$ consists of two pieces: $V_S=V_S^{(1)}+V_S^{(2)}$. 
The first is truly incalculable, as it is quadratically divergent with the cut-off
$V_S^{(1)}= \Lambda^2 M_S^2$. There is no symmetry reason to really control this contribution,
which for $\Lambda\sim M_P$ becomes, understandably,  of the order of the supersymmetry breaking scale in the hidden sector $M_I^4$. The presence of $V_S^{(1)}$ does not only disrupt our dynamically critical minimum, but also implies a mimimum for $M_S$ which is either $0$ or $O(\Lambda)$. In other words $M_S$ is no longer a flat direction.  Without any solid physical motivation one must then assume that by some
clever short-distance conspiracy  $V_S^{(1)}\equiv 0$. Yet this not sufficient, due to the
second contribution
\be
V_S^{(2)}= c_0(M_S) M_S^4 \left[ 1 +O\left( \alpha(M_S)\right) \right] ,
\label{v2}
\ee
where the $O(\alpha)$ term indicates threshold correction effects at the supersymmetric 
scale and where $c_0$ satisfies an RG equation 
\be
\frac{d c_0}{d \ln Q}= \frac{1}{64\pi^2}\left[\frac{\rm Str M^4}{M_S^4}+O\left( \frac{\alpha}{4\pi}\right)
\right] .
\ee
Here $M$ is the mass matrix of all particles that become massive through 
their couplings to $M_S$ including, in particular, the supersymmetric partners of SM fields.
The natural size of $c_0$ is $\sim 1/(4\pi)^2\ln M_P/M_S$, which makes $V_S^{(2)}$ parametrically bigger than $V_{MSSM}^{\rm eff}$ in the region $M_S\sim Q_c$. Then,
in order to preserve the minimum of the full potential $V_{MSSM}^{\rm eff}+V_S^{(2)}$ near the critical point  $M_S\sim Q_c$, also $V_S^{(2)}$ should independently  have a minimum in 
this region. As the stationary points of $V_S^{(2)}$ are determined by dimensional transmutation
through the logarithmic evolution of $c_0$, this coincidence represents a tuning, which we can 
roughly estimate to be of order $1/(\ln  M_P/Q_c)$. Unfortunately, this is precisely what a
dynamical-relaxation model was designed to avoid.
Moreover  the presence of $V_S^{(2)}$ would destroy the prediction $\ln Q_c/M_S= \frac{1}{2}$.
Finally one could argue that, although there is no solid field-theoretic reason to neglect $V_S$, 
perhaps this could follow from whatever mechanism solves the cosmological constant problem. 
We believe it is difficult to imagine how this could work. Indeed from a strict field-theoretic point of view the only distinction between $V_{MSSM}^{\rm eff}$ and $V_S^{(2)}$ is diagrammatic: the latter is 
 determined by 1PI diagrams of supersymmetric SM fields, while the former involves also the  
 one-particle reducible diagrams with tree-level
Higgs exchange. How can the solution of the cosmological constant problem distinguish among different contributions to the potential of the same field $M_S$? Perhaps the only way to proceed is to
 see if there are consequences following from the vanishing of the potential at the minimum. Neglecting $V^{(1)}_S$,
 again without any  explanation, the potential $ V=c_{\rm eff}(M_S)M_S^4$ consists of the addition of
 eq.~(\ref{effective}) and eq.~(\ref{v2}). The coefficient $c_{\rm eff}$ varies logarithmically with $M_S$,
 and it is in principle possible to fine tune the parameters so that $c_{\rm eff}$ and 
 its derivative $c_{\rm eff}^\prime$
 vanish simultaneously at some point. This point would correspond to a minimum with vanishing vacuum energy. It is easy to see that even this criterion in no way singles out the minimum of eq.~(\ref{effective}).

\section{Distribution of Supersymmetry-Breaking Scales}
\label{distrib}
In this section we shall produce an argument on the possible distribution
of the supersymmetry-breaking scale based on simple effective field theory.
Our considerations and results are in line with what was previously found  
in type IIB Calabi-Yau orientifolds \cite{vacua} or, in the same spirit of this section, 
in effective supergravity \cite{Dine:2005yq}. 

Suppose we have a general supersymmetric theory with $N$ chiral
superfields $\Psi_i$, and a general superpotential $W(\Psi)$ and
K\"ahler potential $K(\Psi,\Psi^\dagger)$. We will assume that both of
them include higher-dimension operators suppressed by some
fundamental scale $M_*$, that we set to unity in this discussion. We will 
also ignore supergravity corrections by assuming that $M_*$ is
parametrically smaller than $M_P$; all the vacua that we will
find in our analysis below are then smoothly deformed into vacua of
the full theory with supergravity effects included.

Of course, there will be a large number (exponential in $N$) of
supersymmetric minima associated with the stationary points  of $W$.
However, we also expect to have a large number of metastable
non-supersymmetric minima. It is easy to see that this is only
possible due to higher-order terms in the K\"ahler potential. Indeed, with a
canonical K\"ahler potential, any non-supersymmetric extremum is either a 
saddle point or it is associated with an exactly flat direction. Consider, for
instance, 
the theory of a single chiral superfield with
\begin{equation}
W = a_1 X + \frac{a_3}{3} X^3 + \frac{a_5}{5} X^5 + \cdots .
\end{equation}
For simplicity, to begin with we assume a discrete $R$ symmetry that
makes the superpotential odd in $X$ and makes $K$ a function only of
$X^\dagger X$. For generic coefficients $a_i$, 
supersymmetric minima do exist. There is also a local maximum of the
potential at $X = 0$ with a quadratic instability
along the direction where  $a_1^* a_3X^2$ is real and negative. If we tune
$a_{3,5,..} = 0$, then we have supersymmetry breaking but also an exactly flat
direction for $X$. This is why an arbitrarily small $a_{3,5..}$ can
restore supersymmetry; it lifts the flat direction and drives the field to the supersymmetric
minimum.

However, the story changes if we have corrections to the K\"ahler
potential
\begin{equation}
K = X^\dagger X - \frac{c_2}{4} (X^\dagger X)^2 + \frac{c_4}{9}
(X^\dagger X)^3 + \cdots ,
\end{equation}
because the higher-order terms in $K$ can also lift the flat
direction and, if $c_2$ is large enough relative to $a_3$, stabilize a
non-supersymmetric vacuum at $X = 0$. Indeed, the potential is
\begin{equation}
V = \frac{|a_1 + a_3 X^2 + a_5 X^4 + \cdots|^2}{1 - c_2 X^\dagger X
+ c_4 (X^\dagger X)^2 + \cdots} .
\end{equation}
As long as
\begin{equation}
2 |a_3| < |a_1| c_2 ,
\end{equation}
there is a non-supersymmetric local minimum at $X = 0$. Note that the
condition to find a non-supersymmetric local minimum does not depend on the
values of $a_{5,...}$ and $c_{4,..}$; the reason is that these terms
do not contribute to the quadratic curvature around the extremum at
$X=0$.

Suppose we mediate the supersymmetry breaking to the SM sector via higher-dimensional
operators, so that the overall scale of the soft terms is $M_S \sim |F_X| \sim
|a_1|$ (working in units with $M_* = 1$). For $c_2 \sim O(1)$, to get
a small $M_S$, we need not only $|a_1| \sim M_S$ but also $|a_3| \lsim M_S$
to be small. If the $X$ sector coupled to a landscape of
vacua, so that the complex parameters $a_1,a_3$ scan, it is natural
that, when they are small, they scan with a uniform distribution. So,
since these are two complex parameters, the number of vacua with
supersymmetry breaking scale smaller than $M_S$ is
\begin{equation}
N(M_S) \propto M_S^4.
\end{equation}

Let us now consider the most general case where $X$ is not charged
under any symmetries, so that $W$ and $K$ are general functions of $X$.
By shifting $X$, we can always assume without loss of generality
that there is an extremum for $X$ located at $X=0$, and we can
expand $W,K$ around this point
\begin{equation}
W = \sum_n \frac{a_n}{n} X^{n }, \, \, ~~~~K = \sum_{p,q}
\frac{c_{pq}}{(p+1)(q+1)} X^{(p + 1)} X^{\dagger (q+1)} ,
\end{equation}
with $c_{00}=1$ and $c_{pq}^*=c_{qp}$.
The potential, expanded at quadratic order in $X$, is
\begin{equation}
V=\frac{\left| \partial_X W \right|^2}{\partial_X \partial_{X^\dagger} K}=
| a_1|^2+\left( k_1 X +k_2 X^2 +{\rm h.c.} \right) +k_3 |X|^2+\dots 
%V = \frac{|\sum a_n X^n|^2}{\sum c_{p,q} X^p X^{\dagger q} +
%\mbox{h.c.}}=|a_1|^2+(a_1^*a_3X^2+\mbox{h.c.})+|a_2 X|^2+|a_1|^2c_2|X|^2+\dots
\end{equation}
\bea
k_1&\equiv & a_1^* \left( a_2-a_1 c_{10}\right) \\
k_2&\equiv & a_1^* \left( a_3-a_1 c_{20}\right) -c_{10}k_1 \\
k_3&\equiv & |a_2|^2-|a_1|^2c_{11}-\left( c_{10}^* k_1 +{\rm h.c.} \right) .
\eea
The conditions to have a stable local minimum at $X=0$ with a supersymmetry-breaking
scale equal to $M_S$ are
\beq
|a_1|=M_S,~~~~~k_1=0,~~~~~|k_3|>2|k_2|.
\eeq
For $c_{pq}=O(1)$, these conditions require that $|a_{1,2}|=O(M_S)$ and
$|a_3|\lsim M_S$. If the 3 complex parameters $a_{1,2,3}$ scan uniformly, then we obtain
\begin{equation}
N(M_S) \propto M_S^6.
\end{equation}

Note though that we can get different powers with different
assumptions about the landscape sector. For instance, if it has a CP
symmetry that makes all the parameters real, then we would have
$N(M_S) \propto M_S^3$, while if it also has the discrete $R$ symmetry
of the previous example we would have $N(M_S) \propto M_S^2$.

Of course the non-supersymmetric minima we are considering are
unstable to decaying to a supersymmetric vacuum,  but the lifetime can be exponentially long.
Indeed, we expect the bounce action to scale like $S\sim (\Delta X)^4/\Delta V$ where $\Delta X$
and $\Delta V$ are the shifts in field expectation value and potential energy between the two minima. In our case, $\Delta V=|a_1|^2$ and the location of the closest supersymmetric 
minimum is determined by the quartic term $a_4$ in $W$ and therefore 
$\Delta X \sim |a_1|^{1/3}$.
This gives a lifetime $ \Gamma\sim 
{\rm exp}-(M_*/M_S)^{2/3}$,
which is extremely small as soon as supersymmetry is broken below $M_*$. 

We have phrased the discussion as though the $X$ sector is separate
from the landscape sector, but in fact our conclusions apply to supersymmetry
breaking on a generic supersymmetric landscape. It is clear that in order to
find supersymmetry-breaking extrema, some fields in the theory must become
light; indeed, there must be a massless goldstino. However with a
completely generic superpotential, we expect that all the fields are
heavy with $O(1)$ masses. In some places in field space, though,
it may happen that {\it one} field $X$ is light while the remaining
fields $\phi_i$ are heavy. We can then expand $W$ as
\begin{equation}
W = W_0(\phi) + W_1 (\phi) X + \frac{1}{2} W_2(\phi) X^2 +
\frac{1}{3} W_3(\phi) X^3 + \cdots
\end{equation}
with the remaining fields $\phi$ having an exponentially large
number of supersymmetric minima. In these minima, the parameters in the $X$
theory will scan, and again, when these parameters are small, the
scanning can be taken to be flat.

Note that the scanning for the vacuum energy is completely
independent of the supersymmetry breaking scale~\cite{susd}. In all vacua (supersymmetric
and non-supersymmetric), we will in
general have $\langle W \rangle \equiv a_0\neq 0$. 
When gravity is turned on,
this gives a negative contribution to the cosmological constant
$\Lambda_{SUSY} = - 3 |W|^2/M_P^2$, which is parametrically unrelated to
the scale of supersymmetry breaking. For a uniformly-distributed complex 
parameter $a_0$,
the scanning measure is $d^2 a_0 = |W| d |W| = d \Lambda_{SUSY}$,
{\it i.e.} there is uniform measure on cosmological-constant space. Therefore, the
request of a small cosmological constant does not impose additional constraints
on the statistical distribution of $M_S$.

As we have already mentioned there will also
be an exponentially large number of  supersymmetric
stationary points, where all the landscape fields have generically $O(1)$ masses. What role
can these vacua play in our argument? If these landscape vacua remain exactly supersymmetric even after including the possible infrared dynamics of some hidden gauge group, then they
do not play any role in any selection criteria.
Our universe does not appear to be supersymmetric. 
%On one side these vacua will have a negative
%cosmological constant, in contradiction to observation. On the other, the $\mu$ parameter will be
%generically non-zero and electroweak symmetry will also be unbroken. In other words, an exactly supersymmetric world is ruled out by what we already know about  our universe.
 It is still possible, however, that at these minima some low-energy group with a 
 supersymmetry-breaking infrared dynamics survives. It is natural to expect the distribution of $M_S$ from
 these vacua to be roughly logarithmic  $dN\propto d\ln M_S$~\cite{dinet}, very much like the case of 
 $\Lambda_{\rm QCD}$ considered in sect.~\ref{stat}. If the 
 total number 
 of vacua from this branch were large enough, then it would swamp the distribution of $M_S$
 coming from the branch of local supersymmetry-breaking minima we focussed on so far. 
 In that case, the total distribution of $M_S$ would be essentially $dN\propto d\ln M_S$, and
 the Weak Principle would predict $M_S\ll Q_c$. However the relative weight in vacuum statistics of this branch with dynamical supersymmetry breaking very strongly depends on microphysics
 inputs we do not control. On one side, it is not clear how generic it is that these hidden gauge groups lead to dynamical supersymmetry breaking. Also, there is no universal rationale
 to count  the  supersymmetric versus non-supersymmetric local minima,
 even with a simple landscape model with $N$ chiral fields $\phi_i$ ($i=1,... , N$).
 Assume, for instance, the superpotential is a generic polynomial of degree $M+1$ in $\phi_i$.
 Then the number of classically supersymmetric vacua determined by the equation
 \be
 \partial_i W=0
\label{derW}
 \ee
 scales like $M^N$. The non-supersymmetric stationary points are determined by the equation
 \be
 \partial_j \bigl (\frac{\partial_i W}{\partial_i\partial^k K}\bigr )=0 \,.
 \label{WK}
 \ee
 It is easy to see that, by suitably choosing the K\"ahler potential, there can be more solutions to \eq{WK} than to \eq{derW}.
For instance, take the case of just one superfield $X$ with superpotential and K\"ahler 
potential given by
\beq
W=\exp \left( -\lambda X \right) ,~~~~K=2X^\dagger X -{\rm Si}\left( X^\dagger X \right) ,
\eeq
where $\lambda$ is a coupling constant and ${\rm Si}(x)$ is the sine-integral function, such
that the K\"ahler metric is non-singular and positive definite. This
gives a scalar potential
\beq
V=\frac{\lambda^2 e^{-\lambda( X+X^\dagger )}}{2-\cos (X^\dagger X)},
\eeq
which can reach its supersymmetric vacuum only at $X \to \infty$, but has an
infinite number of non-supersymmetric local minima. 
This result can be generalized to the case of $N$ fields. Therefore, depending on the properties of the K\"ahler potential, there may or there may not be more non-supersymmetric than supersymmetric vacua.

The results of this paper depend on the assumption that the tree-level supersymmetry-breaking
vacua dominate in number over those with dynamical supersymmetry breaking.  It is however remarkable that once this assumption is made the
distribution of supersymmetry-breaking vacua depends on a few universal and basic ingredients.
Our conclusion is that, for a ``generic" theory with a large
$N$ number of fields, there can be a huge number of non-supersymmetric vacua.
In the neighborhood of any one of these vacua, the  breaking of supersymmetry can
be characterized by a single field $X$ getting an $F$-component, and
$M_S$ has a distribution
\begin{equation}
N(M_S) \propto M_S^n,
\end{equation}
where $n$ can run from $2$ to $6$ depending on assumptions on the
structure of the landscape sector, with $n=6$ the most ``generic".

Note that for all $n > 2$, we have a {\it huge} preference for
high-scale supersymmetry breaking; in fact, the tuning it takes to get
low-energy supersymmetry with $m_{Z} \sim M_S$ is much bigger than the
standard hierarchy problem $\sim m_Z^2$. For $n=2$, it is about as
tuned as the usual hierarchy problem, although if we manage to argue
that $M_S$ is a loop factor bigger than $m_Z$, we win in tuning by a
factor $(M_S/m_Z)^2$. However, as we argued in this paper, if $Q_c$ has a maximum,
then the statistical preference for high scale $M_S$ is eliminated by the anthropic prior
that electroweak symmetry be broken (Weak Principle).
The little hierarchy remains as the only detectable signal of an extremely atypical choice
of vacuum, which is dictated by the anthropic prior. 

%If we have such a preference for breaking supersymmetry at a high scale, what
%stops us from going all the way up to the fundamental scale $M_*$? The
%answer is metastability of the non-supersymmetric vacuum. As we said,
%there are always lower supersymmetric vacua that $X$ can tunnel to. The main
%instability is not in fact $X$ tunneling to a supersymmetric minimum with the
%parameters of its potential fixed. Rather, getting the local minimum
%for $X$ requires a special choice for the landscape fields $\phi$, in order to
%tune the parameters $a_{1,2,3}$ 
%to small sizes. For large supersymmetry breaking scales, the action for this
%to happen becomes unsuppressed. Thus, the decay rate for the
%non-supersymmetric minimum is

%\begin{equation}
%\Gamma \sim M_*^4 e^{-(M_*/M_S)^q}
%\end{equation}
%for some $q$, and therefore, metastability puts a cutoff on the
%highest $M_S$
%\begin{equation}
%M_{S}^ {max} \sim \frac{M_*}{q \mbox{log}(M_*/H_0)} .
%\end{equation}
%This is important, because it tells us that the only non-supersymmetric vacua
%that are cosmologically stable will indeed be approximately
%supersymmetric. This means e.g. in coupling to the MSSM, that it is
%at least consistent to imagine that the only thing that scans is the
%overall scale of supersymmetry breaking; we don't have to worry about higher
%derivative operators that would split the MSSM spectrum even with
%only a single source of supersymmetry breaking in $X$.

The main result of our paper  relies on the assumption of softly-broken supersymmetry. For instance if supersymmetry were broken at the cut-off scale,
our minimal scenario, where only the overall value of $M_S$ scans, would
hardly be tenable. In that case, all higher supercovariant  derivative
terms in the action would affect the Higgs potential and there could be plenty of other  vacua where $\langle H\rangle $ is  not controlled by radiative electroweak breaking.
However,
if we have such a preference for breaking supersymmetry at a high scale, what
stops us from going all the way up to the fundamental scale $M_*$? The
answer is metastability of the non-supersymmetric vacuum. As we said, one source
of instability is given by tunneling of $X$ to the closest supersymmetric 
minimum with the parameters of its potential fixed. 
However, the local minimum
for $X$ requires a special choice for the landscape fields $\phi$, in order to
tune the parameters $a_{1,2,3}$  to small sizes. So another, potentially more important,
source of vacuum decay is given by tunneling in $\phi$ space. It is reasonable to expect that the euclidean action for these processes will also be proportional to
an inverse power of $M_S$. Thus, the total decay rate for the
non-supersymmetric minimum is expected to be 
\begin{equation}
\Gamma \sim M_*^4 e^{-(M_*/M_S)^q}
\end{equation}
for some $q$.  For large $M_S$, the decay rate is unsuppressed
 and therefore metastability (corresponding to $\Gamma <H_0^4$, where $H_0$ is the present value of the
 Hubble constant) puts a cutoff on the
highest $M_S$
\begin{equation}
M_{S}^ {max} \sim \frac{M_*}{\left[ \ln (M_*/H_0)\right]^{1/q}} .
\end{equation}
This tells us that the only non-supersymmetric vacua
that are cosmologically stable will indeed be approximately
supersymmetric. In particular, this means that it is
at least consistent to imagine that the only thing that scans is the
overall scale of supersymmetry breaking. We do not have to worry about higher-derivative operators that would effectively make all the ratios of soft terms to scan,
even with
only a single source of supersymmetry breaking in $X$.

 \section{Phenomenological Consequences}
\label{pheno}

The proximity of the critical scale $Q_c$ to the supersymmetric mass
$M_S$ can be empirically tested at collider experiments. When the new-particle
spectrum is known, one will be able to reconstruct the running of the Higgs mass parameters
and observe if the critical condition for electroweak breaking is immediately achieved. However,
even without a complete knowledge of the supersymmetric spectrum, we can obtain, under
certain assumptions on the ratios of soft terms, some predictions on the Higgs and stop masses.

Let us work in the limit of large $\tan\beta$ (which is the most favourable case with respect
to the Higgs mass bound), where the critical scale is determined by the condition
$m_2^2(Q_c)=0$. Since we expect a little hierarchy between the supersymmetric and the weak scale,
in order to accurately compute the relevant physical quantities,
we match to the one Higgs SM at the supersymmetric scale and take into account
the leading RG evolution effects down to the weak scale. As it is convenient and customary, we choose
the geometric average of the physical
stop masses ($M_{\tilde t}^2=M_{{\tilde t}_1} M_{{\tilde t}_2}$) as the matching scale from which to compute the infrared logarithms.
Notice that we do not need to specify a relation between $M_S$ 
and $M_{\tilde t}$ since, as discussed in sect.~\ref{stat}, $M_S$ appears in our equations only through the scheme-independent ratio $Q_c/M_S$.
%Notice that we could pick a convention  where $M_S\equiv M_{\tilde t}$. However 
%as $M_S$ appears anyway in our equations through the scheme independent ratio $M_S/Q_c$ 
%we prefer to keep $M_S$ and $M_{\tilde t}$ formally distinct.

After integrating out all supersymmetric particles and the additional neutral and charged Higgs bosons, the Higgs sector is described
by the familiar SM scalar potential
\beq
V={m^2}|H|^2+\frac{\lambda}{2}|H|^4.
\label{potsm}
\eeq
At the scale $M_{\tilde t}$, the Higgs parameters $m^2$ and $\lambda$ are determined by matching 
the supersymmetric theory with the SM:
\beq
m^2(M_{\tilde t})=m_2^2= -\left [\lambda_t^2\left (m_{\tilde t_L}^2+m_{\tilde t_R}^2+|A_t|^2\right )
-\frac{g_1^2}{5} \left (M_1^2+\mu^2\right )
 -g_2^2 \left (M_2^2+\mu^2\right )\right ] \frac{3L_S}{8\pi^2}
 \label{m2high}
\eeq
%\bea
%K_1&\equiv &\frac{t}{16\pi^2}\left[ 3h_t^2\left( \frac{m_{\tilde Q}^2+m_{\tilde U}^2+A_t^2}{M_S^2}\right)
%-3g^2\frac{M_2^2+\mu^2}{M_S^2}-g^{\prime 2}\frac{M_1^2+\mu^2}{M_S^2}\right] \nonumber \\
%&-&\left. \frac{t^2\alpha_sh_t^2M_3
%(M_3+A_t)}{4\pi^3M_S^2}\right|_{M_S} ,~~~~~~~~~~~~
%t\equiv \ln (Q_c^2/M_S^2) ,
%\eea
\beq
\lambda(M_{\tilde t})=
\left. \frac{g^2+g^{\prime 2}}{4} +\frac{3\lambda_t^4}{16\pi^2}X_t \right|_{M_{\tilde t}} ,
\label{qrtc}
\eeq
where $X_t$ is defined in the appendix and $L_S=\ln Q_c/M_S$.
%Here $m_{\tilde Q}$, $m_{\tilde U}$, $A_t$ are the usual soft terms in the stop sector,
%$M_{1,2,3}$ and $\mu$ are the gaugino and higgsino masses.
%The expression for $K_1$, which follows from the definition of $Q_c$, 
%has been obtained by using an expansion of the leading logarithms,
%keeping only linear terms in $t$, and $\alpha_s h_t^2$ contributions to quadratic terms in $t$.
%This approximation is adequate, because we are interested in the case in which $Q_c$ is
%near $M_S$ ($0<t\ll 1$). Notice that there are no finite (non-logarithmic) corrections to $K_1$,
%since these are all absorbed in the definition of $Q_c$.
 In \eq{qrtc} we have also included a
term which is formally a one-loop correction, but which can be numerically very important 
when the trilinear coupling $A_t$ is large. Consistently with our hypothesis, we can
drop terms suppressed by inverse powers of $\tan\beta$ or proportional to $\mu$. Indeed,
for natural values of the soft parameters of order $M_S$, the higgsino mass is expected to be
of order $\mu ={\cal O}(M_S/\tan\beta )$. 

Next, we renormalize the parameters in \eq{potsm} to the scale of the top mass $m_t$,
and also express the result in terms of the $\overline{\rm MS}$ top Yukawa $h_t$ computed at the top scale
\bea
m^2(m_t)&=&m^2(M_{\tilde t})K_1\\
\lambda(m_t)&=&\frac{g^2+g^{\prime 2}}{4} K_2\\
K_1&\equiv&  1-\frac{3h_t^2t_S}{8\pi^2}\\
K_2&\equiv& K_1^2(1+\delta) \simeq
1-\frac{3h_t^2t_S}{4\pi^2}+\delta 
\\
%\left. 1-\frac{3h_t^2}{8\pi^2}\left\{ t_S+\frac{m_t^2}{m_Z^2}
%\left[ 2t_S+X_t+t_S\left( X_t+t_S\right) \left( \frac{3h_t^2}{16\pi^2}-\frac{4\alpha_s}{\pi}\right)
%\right] \right\} \right|_{m_t},
t_S&\equiv& \ln \frac{M_{\tilde t}}{m_t} .
\eea
Here $K_1$ is just the Higgs wavefunction renormalization due to top loops,
while $\delta$ is the full RG improved top-stop additive correction to the Higgs quartic coupling, given in the appendix.
We have used the SM RG evolution, including only effects from top-Yukawa
and strong interactions. We kept linear terms in $t_S$ and $X_t$, and quadratic terms enhanced
by $m_t^2/m_Z^2$. 

The minimization of the potential in \eq{potsm} allows us to express the Higgs and the Z masses
in terms of $m^2$ and $\lambda$:
$m_h^2=-2m^2$ and $m_Z^2=-m^2(g^2+g^{\prime 2})/(2\lambda)$, where parameters are
evaluated at the scale $m_t$.
Using these equations, we can compute the Higgs mass and $M_{\tilde t}$
in terms of $m_Z$, $L_S$ and the ratios among soft terms:
\bea
m_h^2&=&K_2m_Z^2 
\label{spit1} \\
M_{\tilde t}^2&=&\left| \frac{M_{\tilde t}^2}{m_2^2}\right| \frac{K_2}{2K_1}m_Z^2 .
\label{spit2}
\eea
Here $M_{\tilde t}^2/m_2^2$, with $m_2^2$  given by eq.~(\ref{m2high}), obviously  depends only on ratios of soft terms.
\begin{figure}[t!]
\begin{center}
\includegraphics[width=16cm]{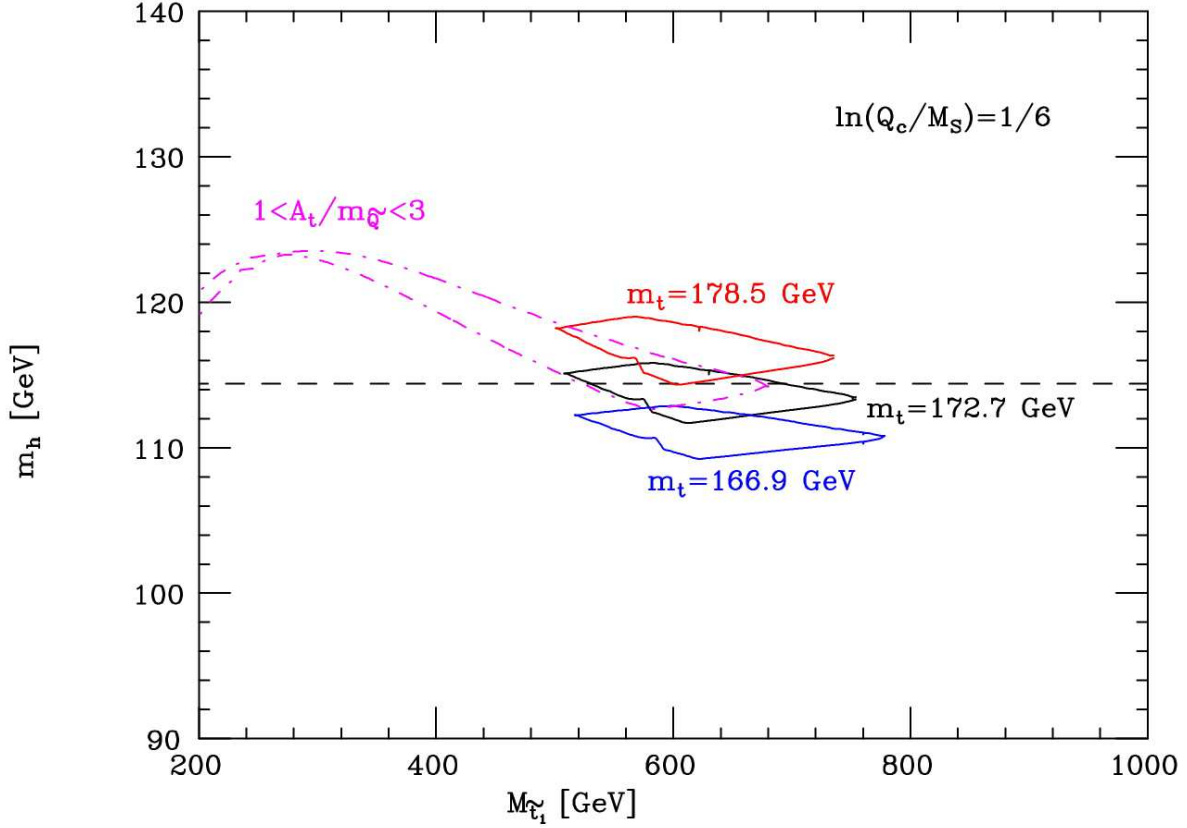}
\end{center}
\caption{
The solid lines are the boundaries of the regions of Higgs mass ($m_h$)
and lightest stop mass ($M_{{\tilde t}_1}$) obtained by 
requiring $\ln Q_c/M_S=1/6$ and by scanning the ratios of soft
parameters in the range $1/2<m^2_{\tilde U}/m^2_{\tilde Q}<2$, 
$0.8<A_t/m_{\tilde Q}<1$,
$1/2<M_3^2/m^2_{\tilde Q}<2$, $1/10<M_{1,2}^2/m_{\tilde Q}^2<1$, under 
the constraint $M_3>200$~GeV, $M_2>100$~GeV, $M_1>
50$~GeV, for the three values of $m_t$ indicated in the figure. The purple
dot-dashed line is the boundary of the analogous region obtained for
$1<A_t/m_{\tilde Q}<3$ and $m_t=172.7$~GeV. The dashed line is the present
lower bound on a SM Higgs-boson mass $m_h>114.4$~GeV.
 }
 \label{fig3}
\end{figure}

In \fig{fig3} we show the values of the Higgs mass $m_h$ and of the lightest stop mass
$M_{{\tilde t}_1}$, obtained from eqs.~(\ref{spit1})--(\ref{spit2}) by fixing $L_S=1/6$ 
(which, as shown in sect.~\ref{distrib}, is the most ``generic" landscape prediction) and by
varying the ratios of low-energy soft parameters
in the range  $1/2<m^2_{\tilde U}/m^2_{\tilde Q}<2$, $0.8<A_t/m_{\tilde Q}<1$,
$1/2<M_3^2/m^2_{\tilde Q}<2$, $1/10<M_{1,2}^2/m_{\tilde Q}^2<1$. These ratios are
varied independently, and therefore we are making no assumption of scalar universality
or gaugino mass unification.
The three regions
shown in \fig{fig3} correspond to the top mass equal to its present central value
$\pm 2\sigma$, and satisfy the requirement $M_3>200$~GeV, $M_2>100$~GeV, $M_1>
50$~GeV. The restricted range of $A_t/m_{\tilde Q}$ is a natural
consequence of the RG running of soft parameters up to a large scale.  Indeed, the gluino mass
gives a large renormalization correction to both parameters, focusing the low-energy
value of this ratio, very much independently of the initial values of the various soft parameters
at the high scale. For instance, taking the top Yukawa corresponding to large $\tan\beta$ and
running up to the GUT scale, we find that $0.79(0.61)<A_t/m_{\tilde Q}<0.97(1.14)$ for
any initial condition of universal scalar masses $m$, of unified gaugino masses $M$ and
trilinear couplings $A$, such that $0<m^2/M^2<1(3)$ and $|A/M|<1(3)$. 

The results in \fig{fig3} show how, for a small value of $\ln Q_c/M_S$, the Higgs mass is
predicted to be very close to its experimental lower bound. On one hand, this can justify why
searches for Higgs and supersymmetric particles have failed so far; on the other hand,
it shows that Higgs and supersymmetric particles lie rather close to their experimental
limits. With large values of $A_t$, heavier Higgs bosons and lighter stops can be obtained,
as illustrated by the region in \fig{fig3} corresponding to a parameter scan in the range
$1<A_t/m_{\tilde Q}<3$ (shown only in the case $m_t=172.7$~GeV). However, as shown above, such large values of this ratio are
unnatural from the point of view of the high-energy theory. Very small values of $A_t/m_{\tilde Q}$
would further lower the prediction for $m_h$, but this also requires a tuning of soft-term
boundary conditions at the high scale.  The prediction on $m_h$ is rather sensitive on the
precise value of $m_t$, as it is well known and as illustrated in \fig{fig3}. A fixed value
of $\ln Q_c/M_S$ also selects a limited range of stop masses. Of course, the precise values
of the allowed $M_{{\tilde t}_1}$
depend on the choice of the interval in which the ratios of soft parameters are varied. The
prediction shown in \fig{fig3} corresponds to the natural hypothesis that these ratios are not
very different from unity. 

\section{Conclusions}

Low-energy supersymmetry still remains the best known candidate to solve the hierarchy
problem, although its natural prediction for new particles with masses around $m_Z$ has not
been confirmed by LEP. The resulting necessity to push $M_S$, the scale of supersymmetric 
particle masses,
almost an order of magnitude above $m_Z$ leads to an apparent fine tuning of few percent
or worse. Although this is a much smaller problem than the original hierarchy, it is still
worrisome for at least two reasons. First, the absence of fine tuning was, after all, the starting
motivation for low-energy supersymmetry. Second, the necessary post-LEP mild tuning puts
in question the chances of discovery at the LHC. Indeed the naturalness criterion,
in spite of its intrinsic arbitrariness, is
necessary to guarantee that supersymmetric particles are accessible to LHC energies, while the more quantitative 
requirement of a thermal relic density appropriate for dark matter is, by itself, not
sufficient. 
 Actually, taking into account LEP bounds and WMAP data, supersymmetric
thermal dark matter requires rather uncharacteristic choices of parameters, raising the
issue of a further source of tuning~\cite{temp}.

Different approaches have been proposed in the literature to reduce the amount of tuning 
or to explain a little hierarchy between $M_S$ and $m_Z$~\cite{lith,other,other2,bs}.  Large trilinear $A$ terms
and a low scale for the original supersymmetry breaking alleviate the problem, but a complete
solution may require a real modification of the minimal supersymmetric SM dynamics at the TeV. Here we
have followed a drastically different approach, appealing to anthropic considerations to 
predict the most probable value of $M_S$, the scale of supersymmetric particle masses.

Symmetry principles have been so successful in particle physics that a general consensus has grown
on the idea that nature is described by a final unique theory, completely determined by 
symmetry properties, possibly allowing no logically consistent modifications. More recently,
this view has been challenged, as a result of both experimental observations and theoretical 
speculations. On one side, the evidence for dark energy reopened the question of the cosmological
constant, which has a satisfactory anthropic justification~\cite{Weinberg}, but no successful explanations based
on symmetry or on dynamics. Also, the negative LEP searches for new physics have created
some conflict in essentially all known models that can {\it naturally} explain the weak scale.
On the theoretical side, the formulation of the string landscape~\cite{lands} together with an
inflationary picture has given a more solid justification for a multiverse description, where some
of the properties of our universe are determined by environmental selection. If true, this description
would represent the ultimate Copernican revolution, since neither the earth nor our observed
universe have a central and unique role in nature.

From a scientific point of view, the great limitation of anthropic considerations, as opposed to
speculations based on symmetry or dynamics, is the dearth of testable predictions. This is
especially true when we cannot directly probe the properties of the statistical
ensemble on which the anthropic principle is applied, as it is the case of the multiverse
picture. Still, it is false that no physical consequences can be obtained. Predictions can be
obtained, although they are different in nature from those derived by dynamics and can 
usually be expressed only in probabilistic terms. A celebrated example is the expectation
that the cosmological constant is of the order of the critical density of our universe~\cite{Weinberg}.
Another use of the anthropic principle is a change of perspective (as, {\it e.g.}, in
Split Supersymmetry~\cite{split}) where arguments based on symmetry properties ({\it
e.g.} the hierarchy problem) are abandoned in favor of mere observational facts. In this 
paper we have offered a new example of an application of the anthropic principle to particle
physics that can lead to testable predictions and we have derived, under certain assumptions, the 
most probable values of supersymmetric particle masses.

First of all, we have recast the hierarchy problem in terms of a criticality condition. Then,
assuming a distribution of vacua where $M_S$ changes and imposing the anthropic
request that electroweak symmetry must be broken by the Higgs field, we have obtained
that $M_S$ is pushed close to $Q_c$, justifying with a statistical argument the 
quasi-criticality of low-energy supersymmetry. In this way we have derived a little hierarchy
between $m_Z$ and $M_S$, {\it a posteriori} explaining why LEP has not discovered
supersymmetry, while maintaining the prediction of discovery at the LHC. We have also
discussed how our conclusions change as we modify the anthropic priors or the number
of scanning parameters.

An interesting conclusion is found when the higgsino mass $\mu$ is allowed to scan 
independently of $M_S$. The anthropic argument shows that values of $\mu$ of the 
order of $M_S$ are preferred, giving a statistical (rather than dynamical) explanation
for the approximate coincidence between the higgsino and gaugino masses. Actually,
for moderate values of $B$, we predict that higgsinos are somewhat lighter and that
$\tan\beta$ is moderately large. Once again, we recall that all predictions
based on the anthropic principle refer to probability distributions. Indeed, we have found
that, for the considered observables, the variance is of the order of the average 
($\langle X^2\rangle \sim \langle X \rangle^2$
for an observable $X$) and therefore large statistical fluctuations are possible. In other
words, our predictions suffer from a ``cosmic variance" problem since, unfortunately,
we can measure only the properties of a single universe, which is actually part of
a large statistical ensemble.

Supersymmetry plays a crucial role in the mechanism we have presented, because it
provides a dynamical explanation for the separation of scales between $Q_c$ and
$M_P$. Here, like in ordinary low-energy supersymmetry, we take advantage of this
natural hierarchy, but we are not trying to derive the absolute value of $M_S$. However,
for a fixed value of the weak scale, we obtain 
a statistical distribution of the relative location of supersymmetry breaking, {\it i.e.}
of $M_S/m_Z$, favoring a little hierarchy. Notice that in this respect, our mechanism could be applied to any
theory with radiative electroweak breaking that predicts a separation between $Q_c$ and
the fundamental high-energy scale.  Our approach is less radical than Split Supersymmetry,
where even the large hierarchy is attributed to anthropic considerations. However,
while in Split Supersymmetry there is no justification for the proximity of the dark-matter
particle mass to the weak scale,  here we retain a dynamical explanation of
this coincidence. 

Our result essentially follows from the observation that electroweak breaking implies
a maximum value of the supersymmetry-breaking scale, $M_S<Q_c$. On the other
hand, the vacuum 
statistics prefer to break supersymmetry at the highest possible scale. Therefore, the
combination of the two effects stabilizes $M_S$ very near the critical value. In other words,
electroweak breaking is a rare phenomenon within the landscape and therefore, once we
impose the prior $\vev{H}\ne 0$, the most likely situation is that $SU(2)\times U(1)$ is only barely
broken, and supersymmetry has to live dangerously close to the critical line of unbroken
symmetry.

We thank Nima Arkani-Hamed for collaboration throughout this project and for numerous suggestions.
We also thank Savas Dimopoulos, Michael Douglas, and Andrea Romanino  for discussions.

\section*{Appendix}

Here we derive some of the equations for the Higgs parameters used in this paper.
First consider the Higgs potential improved by the addition
of the leading top-stop correction $\delta$ to the quartic coupling
%\beq
%V=\frac{g^2+g^{\prime 2}}{8}\Bigl [\left( |H_1|^2-|H_2|^2\right)^2 +\delta |H_2|^4\Bigr ]+m_1^2|H_1|^2
%+m_2^2|H_2|^2-m_3^2\left( H_1H_2+{\rm h.c.}\right) \, .
%\label{1looppot}
%\eeq
\bea
V=&&m_1^2|H_1|^2
+m_2^2|H_2|^2-m_3^2\left( H_1H_2+{\rm h.c.}\right) \nonumber \\
&&+\frac{g^2}{8}\left( H_1^\dagger \vec \sigma H_1 + H_2^\dagger \vec \sigma H_2
\right)^2 +\frac{g^{\prime 2}}{8} \left( |H_1|^2-|H_2|^2\right)^2 +
\frac{(g^2+g^{\prime 2})}{8}\delta | H_2 |^4 ,
\label{1looppot}
\eea
\bea
\delta &=&
\frac{3 h_t^4}{(g^2+g^{\prime 2} )\pi^2 \sin^4\beta}
\left[ t_S+\frac{X_t}{4}+\frac{t_S}{32\pi^2}\left(3 h_t^2-16 g_s^2\right )\left(X_t 
+2t_S \right ) \right] \\
X_t&\equiv& \frac{2\left(A_t -\mu /\tan\beta \right)^2}{M_{\tilde t}^2}
\left[ 1-\frac{\left(A_t-\mu/\tan\beta\right)^2}{12M_{\tilde t}^2}\right] ,~~~~~~~~M_{\tilde t}^2\equiv M_{{\tilde t}_1}
M_{{\tilde t}_2}\\
t_S&\equiv& \ln \frac{M_{\tilde t}}{m_t}.
\eea
Here  $h_t$ is the $\overline{\rm MS}$ top Yukawa at $Q=m_t$ in the SM effective theory.
 
In the presence of a hierarchy between $m_Z$ and the pseudoscalar Higgs mass
$m_A$, the eigenvalues of the mass matrix ${\cal M}^2$ defined by \eq{1looppot} are 
\bea
m^2_{lar}&=& (m_1^2+m_2^2)\left[ 1+O\left( \epsilon_Z \right) \right] \\
m^2_{sma}&=& \frac{m_1^2m_2^2-m_3^4}{m_1^2+m_2^2}
\left[ 1+O\left( \epsilon_Z \right) \right] ,
\eea
where $\epsilon_Z\equiv {m^2_{sma}}/{m_{lar}^2}\sim {m^2_{Z}}/{m_{A}^2}$ will be our expansion parameter. It is convenient to diagonalize ${\cal M}^2$ by redefining
\be
\left (\begin{array}{c}H_1\\ H_2\end{array}\right )=\left (\begin{array}{cc}
\cos\beta'& -\sin\beta'\\
\sin\beta'& \cos\beta'
\end{array}\right ) \left (\begin{array}{c}H\\ H'\end{array}\right ) .
\ee
By integrating out $H'$ in eq.~(\ref{1looppot}) we find the effective potential for $H$
\be
V_{eff}= \frac{m_1^2m_2^2-m_3^4}{m_1^2+m_2^2} |H|^2+\frac{g^2+g^{\prime 2}}{8}\left (\cos^22\beta+\delta\sin^4\beta\right) |H|^4 +O(\epsilon_Z)
\label{effectiveH}
\ee
where, given that $\langle H'\rangle/\langle H\rangle=O(\epsilon_Z)$, the leading result simply amounts to
setting $H'=0$. Notice also that $\tan\beta\equiv \vev{H_2} /\vev{H_1}$ is
equal to $\tan\beta^\prime$
at leading order in $\epsilon_Z$. By minimizing eq.~(\ref{effectiveH}) and expanding ${\cal M}^2$ around its zero at leading order in $L_S$  one obtains eq.~(\ref{derivative}).

We will now instead study the Higgs spectrum for arbitrary $m_A$, but, again, including 
the leading top-stop correction. This corresponds to finding the mass eigenvalues and mixing angles
from the Higgs potential in eq.~(\ref{1looppot}). Defining
\be
\Delta =m_Z^2 \sin^2\beta\,\delta  \qquad\qquad \hat m^2_2=m_2^2+\frac{\Delta}{2} ,
\ee  
the result is
\bea
m_Z^2&=&\frac{2\left( m_1^2-\hat m_2^2 \tan^2\beta\right)}{\tan^2\beta-1}\qquad\qquad 
\sin 2\beta =\frac{2m_3^2}{m_A^2}\\
m_A^2&=&m_1^2+\hat m_2^2 \qquad\qquad m_{H^+}=m_A^2+m_W^2\\
m_{h,H}^2&=&\frac{1}{2}\left\{
m_A^2+m_Z^2+\Delta\pm\sqrt{
\left[ \cos 2 \beta (m_A^2-m_Z^2) +\Delta \right]^2+\sin^22\beta (m_A^2+m_Z^2)^2 } \right\} .
\eea
In the limit $m_A^2\to \infty$ the lightest Higgs mass becomes
\be
m_h^2= m_Z^2(\cos^2 2\beta+\delta \sin^4\beta) ,
\ee
in agreement with eq.~(\ref{effectiveH}).  Notice that only the CP-even Higgs masses are formally affected by the presence of $\Delta$. The tree level relation $m_h^2< {\rm min}(m_A^2,m_Z^2)$ now becomes
\bea
&m_h^2< {\rm min}(m_A^2,m_Z^2+\Delta) \qquad
&{\rm for}\qquad
m_A^2>m_Z^2\frac{\delta}{4+\delta}\\
&m_A^2<m_h^2<m_Z^2+\Delta \qquad
&{\rm for}\qquad
m_A^2<m_Z^2\frac{\delta}{4+\delta} .
\eea
Finally, the square  of the $ZZh$ coupling $\lambda_{ZZh}^2$ is suppressed with respect to the SM value by
a factor
\be
\sin^2(\beta-\alpha)=\frac{1}{2}\left[1+\frac{m_A^2-m_Z^2\cos 4\beta +\Delta \cos 2\beta}
{m_H^2-m_h^2}\right] ,
\ee
while the $ZZH$ and $ZAh$ couplings are proportional to $\cos(\beta-\alpha)$ like in the 
minimal supersymmetric SM. As can be seen from the above equations, in order to have a significant reduction in  $\sin^2(\beta-\alpha)$ one
needs $m_A^2<m_Z^2+\Delta$ as well as $\tan\beta \gg 1$.  However in this region the $ZAh$ coupling is sizeable and moreover, 
for  small $\Delta$,  the threshold 
for $Ah$ production would be significantly below the maximal LEP2 energy. The experimental bound from $Ah$ production then requires a non-negligible $\Delta$ for this region to be viable.
This is the reason why the allowed area in \fig{fig2} does not extend far away from the critical line:
a sizeable top-stop contribution to $\Delta$ is needed.

\end{document}